\documentclass[useAMS]{mn2e}

\usepackage{graphicx}
\def\approxgt{\mathrel{\hbox{\rlap{\lower.55ex \hbox {$\sim$}}
        \kern-.3em \raise.4ex \hbox{$>$}}}}
\def\approxlt{\mathrel{\hbox{\rlap{\lower.55ex \hbox {$\sim$}}
      \kern-.3em \raise.4ex \hbox{$<$}}}}
\def\blfootnote{\xdef\@thefnmark{}\@footnotetext}

\title[The obscuring torus of Markarian~3]{The nature of the torus in the heavily obscured AGN Markarian~3: an X-ray study}

\author[M.~Guainazzi et al.]{
M.~Guainazzi$^{1,2}$\thanks{E-mail:mguainaz@astro.isas.jaxa.jp},
G.~Risaliti$^{3}$,
H.~Awaki$^{4}$,
P.~Arevalo$^{5}$,
F.E.~Bauer$^{6,7,8}$,
S.~Bianchi$^{9}$,
\newauthor
S.E.~Boggs$^{10}$,
W.N.~Brandt$^{11,12,13}$,
M.~Brightman$^{14}$,
F.E.~Christensen$^{15}$,
W.W.~Craig$^{10,16}$,
\newauthor
K.~Forster$^{14}$,
C.J.~Hailey$^{17}$,
F.~Harrison$^{14}$,
M.~Koss$^{18}$,
A.~Longinotti$^{19}$,
C.~Markwardt$^{20}$,
\newauthor
A.~Marinucci$^{9}$,
G.~Matt$^{9}$,
C.S.~Reynolds$^{21}$,
C.~Ricci$^{6}$,
D.~Stern$^{22}$,
J.~Svoboda$^{23}$,
\newauthor
D.~Walton$^{22}$,
W.~Zhang$^{20}$ \\
$^{1}$Institute of Space and Astronatical Science (JAXA), 3-1-1 Yoshinodai, Sagamihara, Kanagawa, 252-5252, Japan \\
$^{2}$European Space Astronomy Center of ESA, P.O.Box 78, Villanueva de la Ca\~nada, E-28691 Madrid, Spain \\
$^{3}$INAF-Osservatorio di Arcetri, Largo E. Fermi 5, I-50125, Firenze, Italy \\
$^{4}$Department of Physics, Ehime University, Matsuyama 790-8577, Japan \\
$^{5}$Instituto de F\'isica y Astronom\'ia, Facultad de Ciencias, Universidad de Valpara\'iso, Gran Bretana N 1111, Playa Ancha, Valpara\'iso, Chile \\
$^{6}$Instituto de Astrof\'{\i}sica, Facultad de F\'{i}sica, Pontificia Universidad Cat\'{o}lica de Chile, Casilla 306, Santiago 22, Chile \\
$^{7}$Millennium Institute of Astrophysics (MAS), Nuncio Monse\~{n}or S\'{o}tero Sanz 100, Providencia, Santiago, Chile \\
$^{8}$Space Science Institute, 4750 Walnut Street, Suite 205, Boulder, Colorado 80301 
$^{9}$Dipartimento di Matematica e Fisica, Università degli Studi Roma Tre, via della Vasca Navale 84, 00146 Roma, Italy \\
$^{10}$Space Sciences Laboratory, University of California, Berkeley, CA 94720, USA \\
$^{11}$Department of Astronomy and Astrophysics, The Pennsylvania State University, 525 Davey Lab, University Park, PA 16802, USA \\
$^{12}$Institute for Gravitation and the Cosmos, The Pennsylvanya State University, University Park, PA 16802, USA \\
$^{13}$Department of Physics, 104 Davey Lab, The Pennsylvania State University, University Park, PA 16802, USA \\
$^{14}$Cahill Center for Astrophysics, California Institute of Technology, 1216 East California Boulevard, Pasadena, CA 91125, USA \\
$^{15}$DTU Space - National Space Institute, Technical University of Denmark, Elektrovej 327, 2800 Lyngby, Denmark \\
$^{16}$Lawrence Livermore National Laboratory, Livermore, CA, 94550, USA \\
$^{17}$Columbia Astrophysics Laboratory, Columbia University, New York, NY 10027, USA \\
$^{18}$Institute for Astronomy, Department of Physics, ETH Zurich, Wolfgang-Pauli-Strasse 27, CH-8093 Zurich, Switzerland \\
$^{19}$Catedr\'atica CONACYT - Instituto Nacional de Astrof\'isica, \'Optica y Electr\'onica, Luis E. Erro 1, Tonantzintla, \\ Puebla, C.P. 72840, M\'exico\\
$^{20}$NASA Goddard Space Flight Center, Greenbelt, MD 20771, USA \\
$^{21}$Department of Astronomy, University of Maryland, College Park, MD 20742, USA \\
$^{22}$NASA Jet Propulsion Laboratory, 4800 Oak Grove Dr, Pasadena, CA 91109, USA; California Institute of Technology, \\ 1200 East California Boulevard, Pasadena, CA 91125, USA \\
$^{23}$Astronomical Institute, Academy of Sciences, Boční II 1401, 14100, Prague, Czech Republic \\
}

\begin{document}

\date{}

\pagerange{\pageref{firstpage}--\pageref{lastpage}} \pubyear{2015}

\maketitle

\label{firstpage}

\begin{abstract}
  In this paper
  we report the results of an X-ray monitoring campaign
  on the heavily obscured Seyfert galaxy Markarian~3
  carried out between the fall of 2014 and the spring of 2015
  with {\it NuSTAR}, {\it Suzaku} and {\it XMM-Newton}.
  The hard X-ray spectrum of Markarian~3 is variable on
  all the time scales probed by our campaign, down to a few days.
  The observed continuum variability is due to an
  intrinsically variable primary continuum seen
  in transmission through a large, but still Compton-thin column density
  ($N_{\rm H}$$\sim$0.8-1.1$\times 10^{24}$~cm$^{-2}$).
  If arranged in a spherical-toroidal geometry, 
  the Compton scattering matter has an opening angle $\simeq$66$^{\circ}$,
  and is seen
  at a grazing angle through its upper rim
  (inclination angle $\simeq$70$^{\circ}$).
  We report a possible occultation event during the
  2014 campaign. If
  the torus is constituted by a system of clouds sharing the same
  column density, this event allows us to constrain their number
  ($17 \pm 5$) and individual column density,
  [$\simeq$$(4.9 \pm 1.5) \times 10^{22}$~cm$^{-2}$].
  The comparison of IR and X-ray spectroscopic results with
  state-of-the art ``torus'' models
  suggests that at least two thirds
  of the X-ray obscuring gas volume might be located within the dust
  sublimation radius.
  We report also the discovery of an ionized absorber, characterized
  by variable resonant absorption lines due to He- and H-like iron.
  This discovery lends support to the idea that
  moderate column density absorbers could be due to clouds evaporated
  at the outer surface of the torus, possibly
  accelerated by the radiation pressure
  due to the central AGN emission leaking through
  the patchy absorber.
  
\end{abstract}

\begin{keywords}
Galaxies:~Seyferts -- Galaxies:~active -- X-rays:~galaxies -- X-rays:individual:~Markarian~3
\end{keywords}

\section{Introduction}

\noindent
    
A breakthrough in our understanding of the physical properties and geometrical configuration
of gas and dust in the innermost parsec around accreting super-massive black holes came recently
from dust reverberation experiments in the Near InfraRed (NIR). By measuring the
response of hot dust emission to changes in the ionizing continuum,
it has been possible to determine the typical size of the innermost boundary of the
dusty ``torus'' in about 20 nearby Active Galactic Nuclei
(AGN) \cite{suganuma06,kishimoto07,koshida14,pozonunez14,jun15}.
Coupled with imaging NIR and mid-IR interferometry \cite{jaffe04,burtscher13} and models of ``clumpy
tori'' \cite{nenkova08}, these observational results
set unprecedented constraints on the structure and location of the azimuthally symmetric
obscuring material invoked by ``unification-by-orientation'' AGN scenarios
\cite{antonucci85,antonucci93,netzer15}. The most important collective result of
these experiments is the discovery of a scaling law between
the spatial scale of the innermost region
of the host dust and the continuum luminosity. Whenever measurements in the
NIR are available together with optical reverberation mapping results, the
hot dust spatial scale is larger
by a factor of about  three
than the optical Broad Line Region (BLR). This vindicates
the old idea that the outermost region of the BLR is set by
the dust sublimation radius \cite{netzer93}.

However, IR measurements probe only the dusty phase of the obscuring matter in
the AGN environment. In order to obtain the full picture, X-ray
measurements are
required. X-rays probe emission
(and therefore its obscuration) along the line-of-sight to the central engine
down to scales on the order of a few gravitational radii from the event horizon
\cite{risaliti05,chartas09,morgan12}.
Serendipitous X-ray occultation measurements demonstrate that highly dynamical
systems of
X-ray obscuring clouds are present at all scales from the innermost region
of the BLRs to the torus \cite{bianchi12,markowitz14,torricelliciamponi14}.

In 2012, Guainazzi et al. (2012; G12) analyzed all the {\it XMM-Newton}, {\it Suzaku}
and {\it Swift} X-ray observations of the nearby ($z$=0.014)
AGN Markarian~3 performed in
this century.
Known since the dawn of X-ray astronomy, Markarian~3 hosts a
heavily obscured AGN \cite{cappi99} with a Seyfert~2 optical
classification \cite{khachikian74}.
The most prominent feature in its
X-ray spectrum is a strong iron
K$_{\alpha}$ fluorescent line \cite{awaki91}.
The relation between
the Fe K$_{\alpha}$ Equivalent Width ($\simeq$1000~eV) and the ratio
between the X-ray and the Balmer-corrected [OIII] flux ($\sim$0.14)
is consistent with a heavily obscured Compton-thin, or with a Compton-thick AGN
\cite{bassani99}\footnote{Following the usual convention, we assume
$N_{H,th}$$\equiv$$\sigma_t^{-1} \simeq 1.5 \times 10^{24}$~cm$^{-2}$ as the
column density threshold separating Compton-thin from Compton-thick AGN,
where
$\sigma_t$ is the Thompson scattering cross-section.}.
While the soft X-ray
emission stayed constant (within the flux cross-calibration uncertainties
among different observatories) over the last 15 years (as well as during a prior
13~year campaign discussed by Iwasawa et al. 1994), the hard X-ray flux above
4~keV showed significant variability down to time-scales as short as two months.
Moreover, assuming that the light curves in the 3--5~keV and
15--150~keV energy bands are correlated, the former lagged the
latter with a minimum measured delay of $\approxgt$1200~days (G12).
These experimental results together with the discovery of extended hard
($E$$\approxgt$3~keV)
X-ray emission on scales of $\simeq$300~pc suggested that
the X-ray emission in Markarian~3 could directly
probe the clumpy nature of the gas and
dusty phase of the obscuring matter in AGN.

However, the results published in G12 were primarily based on spectra taken at
energies lower than 10~keV. This affected their results in two ways: a) it did not
allow
them to probe the line-of-sight column density simultaneously with the
spectral component associated with
Compton scattering of the global distribution of
clouds surrounding the AGN; b) it did not fully remove
uncertainties regarding the uniqueness
of the spectral decomposition, due to the expected spectral complexity in the
energy range where the analysis was possible. Moreover, the observations of
Markarian~3 taken before 2012 covered a sparse time pattern,
inadequate to determine variability time-scales.

Recently, Yaqoob et al. (2015) challenged G12's interpretation of the
X-ray light curve of Markarian~3 on different astrophysical grounds.
They showed that the G12 interpretation
was primarily driven by using models for the
reflection continuum with an infinite matter column density. Using
models that fit explicitly the global column density of the reflector
and treat self-consistently the reprocessing of the X-ray continuum and
the X-ray fluorescence, Yaqoob et al. (2015) found a line-of-sight
column density significantly lower than the Compton-thick
limit. This result
falsifies the basic assumption in G12, {\it i.e.} that the 4--5~keV
energy band can be used to probe the reprocessing matter.

In order to overcome
the sparse sampling of the existing archival data,
an observational campaign on Markarian~3 was
designed and carried out between autumn 2014 and spring 2015, based
on {\it NuSTAR} \cite{harrison13}. Thanks to its hard X-ray
focusing optics, {\it NuSTAR} allows a giant leap in sensitivity above 10~keV
when compared to any prior X-ray observatories.
In this paper we present the main results of this campaign, characterizing
the properties of the obscuring gas and dust surrounding the nuclear environment
of this archetypical obscured AGN. Besides regular monitoring
of Markarian~3 with {\it NuSTAR}, the campaign included
quasi-simultaneous observations with the CCD detectors
on board {\it Suzaku} and {\it XMM-Newton}, providing better energy resolution
at the energy of the iron atomic transitions. We also
present in this paper a reanalysis
of the 2005 {\it Suzaku} data originally discussed by Awaki et al. (2008), as well
as archival {\it Chandra}/HETG data at the highest available
spectral resolution in
the the iron energy band.

The paper is organized as follows: we describe the observational campaign in
Sect.~\ref{sect_campaign}. The data reduction and spectral analysis are
described in Sect.~\ref{sect_reduction}. The results of
the spectral analysis are presented
in Sect.~\ref{sect_dataanalysis}. We discuss our findings in Sect.~\ref{sect_discussion},
and summarize them in Sect.~\ref{sect_summary}.
The following cosmological
parameters were used to calculate luminosities:
$H_0$=70~km~s$^{-1}$~Mpc$^{-1}$, $\Lambda_0$=0.73,
and $\Omega_M$=0.27 \cite{bennett03} to ease comparison with G12.
With this choice, 1'' corresponds to 270~pc at the distance of Markarian~3.

\section{The observational campaign}
\label{sect_campaign}

The {\it NuSTAR} monitoring campaign on Markarian~3 was divided into two seasons.
A first series of five observations was
performed between 2014 September 7 and
October 23. The time separation between consecutive observations ranges
between 7 and 22 days. Prompted by the discovery of variability between
observations separated by the shortest time interval, a second observational
campaign was organized for Winter-Spring 2015. Four additional
observations were performed
between 2015 March 19 and April 8, with a minimum separation of three days between
consecutive observations.
Table~\ref{tab_log}
presents a log of the observations discussed in this paper.
\begin{table*}
\caption{Log of the observations discussed in this paper. In the 5th column: ``S''={\it Suzaku}; ``X''={\it XMM-Newton}.}
\label{tab_log}
\begin{tabular}{lcclcc} \hline \hline
Obs.\# & Start time & $T_{exp}$ & Obs.\# & Start time & $T_{exp}$ \\
 &  & (ks) &  &  & (ks) \\
\hline
\multicolumn{3}{l}{{\it NuSTAR}} & \multicolumn{3}{l}{Quasi-simultaneous with ...} \\
60002048002 & 2014-09-07T17:26:07 & 30.0 & ... & ... & ...  \\ 
60002048004 & 2014-09-14T10:51:07 & 33.5 & ... & ... & ...  \\ 
60002048006 & 2014-10-01T12:41:07 & 33.2 & 709022010 (S) & 2014-10-01T22:08:01 & 20.6 \\ 
60002048008 & 2014-10-09T04:36:07 & 26.5 & ... & ... & ...  \\ 
60002048010 & 2014-10-23T06:01:07 & 30.9 & 709022030 (S) & 2014-10-23T03:37:31 & 20.3 \\ 
60002049002 & 2015-03-19T12:46:07 & 23.2 & 0741050101(X) & 2015-03-19T17:55:25 & 3.1 \\ 
60002048012 & 2015-03-23T05:31:07 & 26.7 & 709022040 (S) & 2015-03-23T05:48:56 & 18.7 \\ 
60002049004 & 2015-04-05T01:31:07 & 24.7 & ... & ... & ... \\ 
60002049006 & 2015-04-08T08:36:07 & 25.2 & 0741050201 (X) & 2015-04-08T16:46:38 & 3.8\\ 
\hline \hline
\end{tabular} 
\end{table*}

Whenever possible, {\it Suzaku} or {\it XMM-Newton} observations were
planned quasi-simultaneously
with {\it NuSTAR}. This was possible on 2014 October 1,
October 23 with {\it Suzaku} and
2015 March 23, and on 2015 March 19 and April 8 with
{\it XMM-Newton}. The {\it XMM-Newton} observations
were performed at the end of the 48-hour spacecraft orbit, close to the
radiation belts. The exposure times were therefore shorter than
1 hour, compared to the 4 to 7 hours of the {\it NuSTAR} and {\it Suzaku}
observations.
Despite the low signal-to-noise, the inclusion of the EPIC-pn spectrum
in the fit improves the precision in the determination of the Fe K$_{\alpha}$
centroid energy by $\simeq$25\% in the April 8 observation
(from $\Delta E$$\simeq$40~eV to $\Delta E$$\simeq$30~eV), thanks to the better
energy resolution of the CCD cameras.
An additional {\it Suzaku} observation was
performed on 2014 October 7. As this observation does not overlap with
the closest {\it NuSTAR} observation (October 9),
these data will not be discussed
in this paper.

In this paper, spectra extracted from overlapping observations are
analyzed together. To maximize signal-to-noise
we did not attempt to find common
Good Time Intervals between overlapping observations.
Spectral fits are typically dominated by {\it NuSTAR} due to the significantly
larger number of collected photons. The CCD measurements provide primarily
an independent check on the accuracy of the energy scale at the
Fe emission line energies. Given the typical time-scales
and spectral dynamical range, not using strictly simultaneous observations
could in principle induce systematic
uncertainties in the derived spectral parameters.
However, we estimate that this uncertainty is smaller than
the statistical errors.

\section{Data reduction}
\label{sect_reduction}

\subsection{{\it NuSTAR}}

\begin{figure}
  \includegraphics[height=85mm,angle=-90]{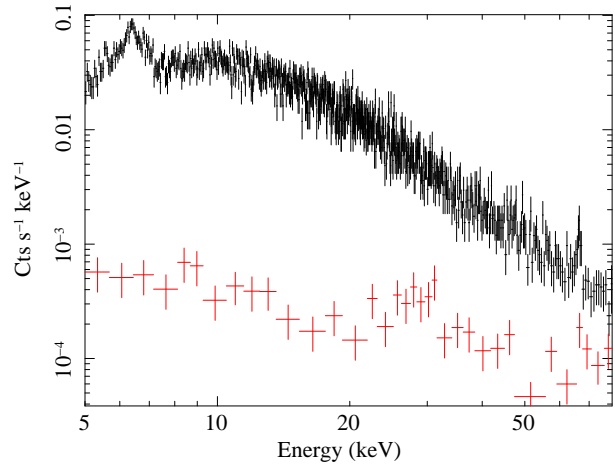}
  \caption{Comparison between the source plus background ({\it top; black})
    and the background {\it NuSTAR} spectrum ({\it bottom; red}) for
    Obs.\#60002049002, corresponding to the lowest {\it NuSTAR}
    count rate measured during
    the Markarian~3 2014--2015 campaign.}
  \label{fig_nustarsrvsbk}
\end{figure}

{\it NuSTAR} data were reduced using the software suite available in {\sc Heasoft}
version 6.16, using the CALibration DataBase version 1.1 (2015 March 16).
Calibrated event lists were produced with {\tt nupipeline} version 0.43.
Light curves and spectra were generated with {\tt nuproducts} 0.2.8,
using standard data selection criteria.
Observations with {\it Chandra} ACIS unveiled extended emission
$\pm$2\arcsec\, in size along a E-W axis aligned with
the E-W optical Narrow Line Regions in the soft X-ray band
\cite{sako00} with fainter emission extending up to 15\arcsec\, in radius.
The iron line emission could also be similarly
extended (G12). This means that the X-ray emission of
Markarian~3 is basically point-like for the
{\it NuSTAR} telescopes' Point Spread
Function (60\arcsec\, half-power diameter, Madsen et al. 2015).
We therefore extracted
source plus background photons from a 122\arcsec\, radius
circular regions centered on the
nominal optical coordinates of the AGN
($\alpha_{2000}$=6:15:36.316, $\delta_{2000}$=+71:02:12.51; J2000).
Besides the emission from the Markarian~3 AGN, the
{\it Chandra} ACIS image shows three additional point sources in
the NuSTAR aperture.
These sources are not detected above the background fluctuations in
a $E > 4$~keV ACIS-S image of the Markarian~3 field taken in 2013,
January 13 (when the image of Markarian~3 comprised about 2000 counts).
Their distances from the Markarian 3 nucleus are between 50\arcsec and
110\arcsec. They would be detected at least as a distortion of the NuSTAR
PSF, or as a displacement of its centroid
if they would significantly contribute to the observed flux during the
NuSTAR observations. We estimate that their contribution to the flux above 4~keV
in the NuSTAR aperture
is $\ll$1\% during the monitoring campaign discussed in this paper.
Background
spectra were extracted from 84\arcsec\,
circular regions around the sky coordinates ($\alpha_{2000}$=6:16:31.307,
$\delta_{2000}$=+71:02:54.05), after checking that this
region is free
from serendipitous contaminating sources in the 3-80~keV {\it NuSTAR} images.
The total counts in the background regions in the different observations
are consistent with pure Poissonian distributions whose mean, $\mu$,
and variance, $\sigma^2$, agree within 0.3\%. No spatial
gradient is seen in the background images. The fraction of the background
counts with respect to the source plus background counts is $\simeq$1\%
at 6~keV, and $\simeq$10\% at 50 keV (Fig.~\ref{fig_nustarsrvsbk}).

\begin{figure}
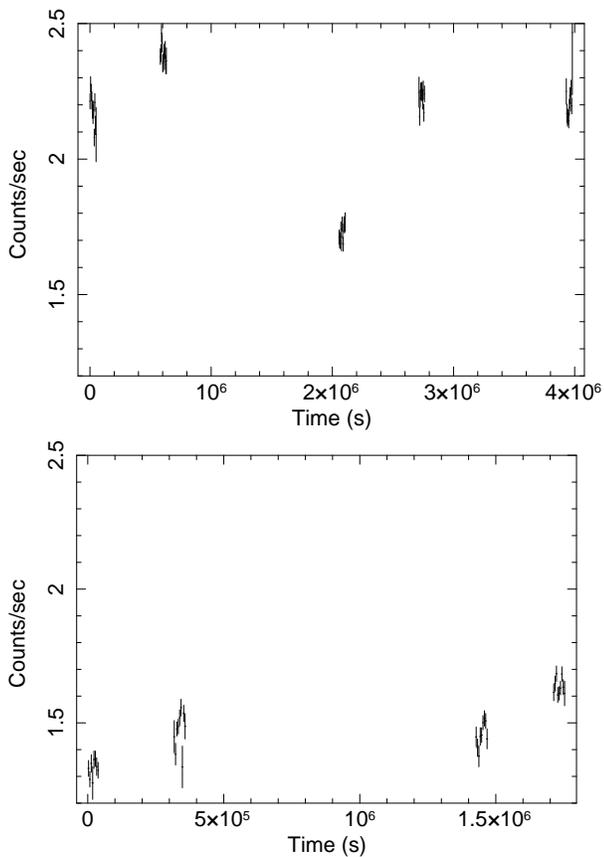

  \includegraphics[height=85mm,angle=-90]{Markarian3_NuSTAR_2014_lc.eps}
  \includegraphics[height=80mm,angle=-90]{Markarian3_NuSTAR_2015_lc.eps}
  \caption{3--50~keV {\it NuSTAR} light curve of Markarian~3 during the 2014
    ({\it upper}) and 2015 ({\it lower})
    observational campaigns. The binning time is 5000~s. The light curves are
    shown on the same y-axis scale to ease comparison. Error bars represent
  1-$\sigma$ level uncertainties.
  }
  \label{fig_NuSTARlc}
\end{figure}

The {\it NuSTAR} 3--50~keV light curve shows large-amplitude variability with
a dynamical range larger than a factor of 2 (Fig.~\ref{fig_NuSTARlc}).
The largest flux changes
($\simeq$70\%) occurred between the second (2014 September 14) and the
3rd (2014 October 1) epochs.
We extracted the light curves from each
individual observation in the 3--7~keV,
7--15~keV, and 15--50~keV energy bands to check for any significant
inter-observation
flux variability. As
a crude variability estimate, we calculated the reduced $\chi^2$ when
fitting each light curve with a constant. In only four light curves is
$\chi^2$ larger than the 1/$N_{lc}$ percentage point, where $N_{lc}$=27
is the total number of light curves (3 energy ranges times 9 observations):
the 7--15~keV light curve on 2015 March 19,
and the 15--50~keV light curves on 2014 September 7, 2015 March 19, and
2015 April 5. In all cases the comparatively high values of $\chi^2$ are
dominated by individual bins with a low effective exposure. We therefore
conclude that inter-observation variability does not effect the
results discussed in this paper.

\subsection{{\it Suzaku}}

We followed the {\it Suzaku} data reduction procedure as described in G12.
Briefly, we used the {\it Suzaku} data analysis software
included
in {\sc Heasoft} version 6.16, and CALDB version 1.1 (2014 October 10).
Data were reduced, and spectral products and responses were extracted
following the same procedures as in Awaki et al. (2008). 
Due to spacecraft battery re-charging issues, the Hard X-ray Detector (HXD)
was switched off during all the
observations of the Markarian~3 monitoring campaign.
Only data obtained with the CCD cameras on-board {\it Suzaku} (X-ray
Imaging Spectrometer, XIS) \cite{koyama07} are therefore
discussed in this paper.
Source plus background spectra were extracted from circular regions centered
on the X-ray source centroid, using a radius between about
190'' and 240'' to optimize the signal-to-noise of each individual
observation. Background spectra were extracted from circular regions on
the same camera chip,
avoiding the region in detector coordinates illuminated by the
calibration source, as well as serendipitous contaminating sources in the
XIS field-of-view -- in particular, IXO-30 \cite{bianchi05b,pounds05}.
We merged together the spectra of the operational
front-illuminated CCDs (XIS0 and XIS3; ``FI'' chips hereafter),
but fit separately the spectra of
the front- and back-illuminated units due to their different responses.

\subsection{{\it XMM-Newton}}

Also for the {\it XMM-Newton} data we followed the same data reduction procedure
as in G12. Because we are interested in the variability
properties of Markarian~3 in the hard X-ray band ({\it i.e.}, above
4~keV), we do not discuss the high-resolution
Reflection Grating Spectrometer (RGS) data in this paper. Furthermore, we
consider only spectra extracted with the CCD EPIC-pn camera
\cite{struder01}, due to its higher effective area.
Calibrated event lists, spectra and responses were generated with SAS version
14 \cite{gabriel03}, using the most updated calibration files available
at the time the data reduction was performed (April 2015). Source plus
background spectra were extracted from circular regions of 30'' radius.
Background spectra were generated from circular regions of 60'' radius,
extracted from the same CCD and at the same height in detector coordinates
as the source to ensure that the same Charge Transfer Inefficiency correction
applies, because this correction depends on the distance from
the readout node. 

\subsection{Spectra handling}

In fitting the spectra, we employed the Cash
goodness-of-fit statistic \cite{cash76},
that is the appropriate maximum likelihood for
Poissonian data.
More specifically, we employed the W-statistics implemented in {\sc Xspec}
v12.8.2 \cite{arnaud96} through the command {\tt statistic cstat}
when source plus background and background spectra are used. In
order for the algorithm to work properly, we rebinned the source
plus background spectra to ensure that each spectral channel has at least
one count (K.Arnaud, private communication).
We did not attempt at modeling the background spectra,
due to their complexity and not fully understood time-variability. We
used instead the approach, standard in X-ray astronomy, of subtracting
from the source plus background spectrum an
appropriately rescaled background spectrum
prior to applying the forward-folding spectral
fitting algorithm.
Unless otherwise specified, statistical uncertainties are quoted
hereafter at the
90\% confidence level for one interesting parameter \cite{lampton76}.
We fit the
CCD spectra in the 4--10~keV energy range, while the {\it NuSTAR}
spectra are fit in the 5--79~keV energy range, where the response is
well calibrated \cite{madsen15}.

The spectra analyzed in this paper were extracted
from sky regions sharing the same centroid, but whose size was optimized
to obtain the best signal-to-noise for each instrument. The difference in
size could in principle introduce spurious variability when comparing
observations taken with different instruments at different epochs. However,
we are confident that this effect is negligible on the following grounds:
a) at all epochs at least a pair of NuSTAR spectra were analyzed.
These spectra were extracted from the same regions in sky coordinates.
They dominate the statistics of the spectral fits. Fluxes and luminosities
in this paper are calculated from the NuSTAR best-fit models; b) models
fitting the {\it Suzaku}/XIS and XMM-Newton/EPIC spectra were multiplied
by constant factors to take into account possible differences in the
instrumental cross-calibration. These factors are always $\le 10\%$,
consistent with known calibration uncertainties \cite{ishida11,madsen15};
c) while the hard X-ray emission of Markarian~3 is extended,
the extension is of a few arc-seconds at most, and its integrated
flux is
only a few percent of the total flux (G12). For instruments with a moderate
spatial resolution as those discussed in this paper, the hard X-ray emission
of Markarian~3 is basically point-like. The measured counts
are therefore properly corrected for the
encircled energy fraction through the effective area when deriving
intrinsic fluxes and luminosities.

\section{Data analysis}
\label{sect_dataanalysis}

\begin{figure}
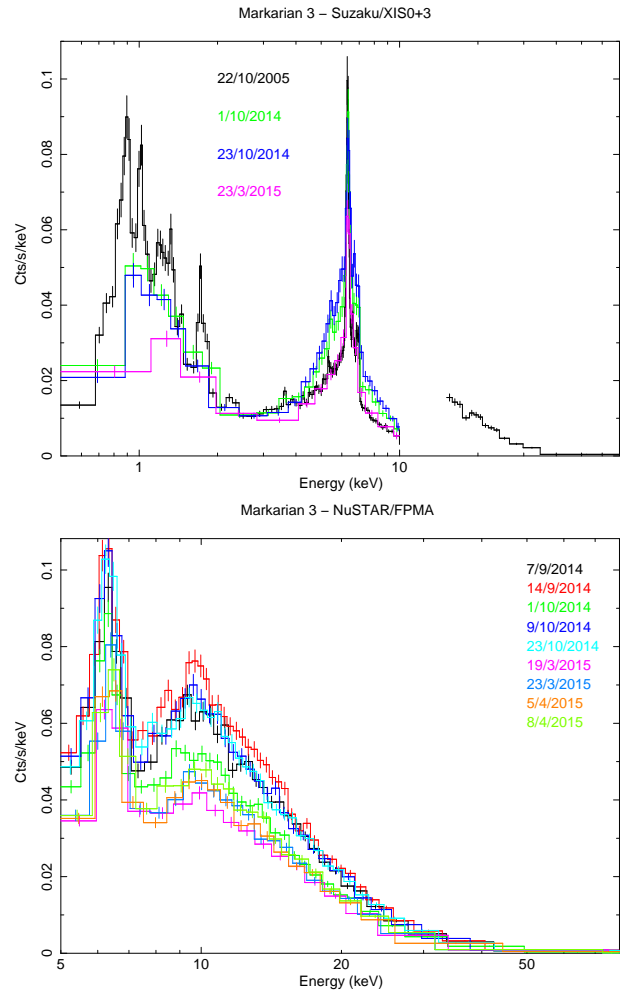

  \includegraphics[height=85mm,angle=-90]{fi_all.eps}
  \includegraphics[height=85mm,angle=-90]{s_FPMA.eps}
  \caption{
  {\it Suzaku} FI-CCD ({\it top}) and {\it NuSTAR} ({\it bottom}) count
  spectra obtained during the 2014-2015 monitoring campaign. The
  {\it labels} indicate the observation start day. The {\it Suzaku}
  plot includes also the 2005 XIS and HXD observation.
  In this Figure the spectra are
  rebinned for clarity to a minimum signal-to-noise ratio of 10
  per spectral channel.
  }
  \label{fig_spectra}
\end{figure}

Fig.~\ref{fig_spectra} shows the {\it NuSTAR} and {\it Suzaku}/XIS
(FI chips only)
spectra taken during the 2014-2015 monitoring campaign. Due to the
stability of the instrumental responses \cite{koyama07,madsen15} these
count spectra can be directly compared to gauge qualitatively the
hard variability pattern above 3~keV, where the
XIS spectra show a dramatic count rate increase due to the
emergence of the obscured AGN component\footnote{The {\it Suzaku}/XIS
is known to be affected of a strongly time-variable
molecular contamination layer. This layer, however, affects the response
only below $\simeq$1~keV.}.

The {\it NuSTAR} spectra in Fig.~\ref{fig_spectra}
show a slight increase of the overall flux between September 7 and September 14
was followed by a sharp $\simeq$70\% flux decrease in the following two weeks.
The 2015 observations exhibit a lower flux than the 2014 ones.
The lowest flux level was measured on March 19, followed by a slow gradual
20\% flux recovery until the last observation on April 8. 
Significant flux ($7.7 \pm 0.1\%$)
and spectral variability is observed down to
the shortest time-scale probed by the monitoring campaign, {\it i.e.}
the four days
between April 19 and April 23. It is the first time that hard X-ray
variability on such a short time scale is reported in Markarian~3.
Prior studies estimated an upper limit of the hard X-ray
variability time-scale of about two months (Iwasawa et al. 1994; G12).

The comparison among the {\it Suzaku} FI-XIS count spectra allows us to
better disentangle the variability pattern of the Fe K$_{\alpha}$ line
and of the underlying continuum due to the better resolution of the CCD cameras at
$\simeq$6~keV. The
spectrum measured during the flux minimum on March 23 ({\it Suzaku} did
not observe Markarian~3 on March 19) exactly overlaps with that
measured during the deep {\it Suzaku} observation in 2005 \cite{awaki08}.
This suggests that such a low flux state corresponds to a
high Compton-reflection versus primary flux fraction
due to the expected stability of this spectral component on the
time-scales probed by the 2014--2015 monitoring campaign (and confirmed
{\it a posteriori} by the results discussed in this paper). 
The comparison between the {\it NuSTAR} March 19 and 23 spectra further suggests
that the minimum flux state measured by {\it Suzaku} was preceded by a
state with an even higher reflection fraction.

Even in the restricted
energy decade discussed in this paper, the Markarian~3 spectrum
is complex. In order to
understand the contribution of each of these components to the observed
variability pattern, and to elucidate the origin of the variability, the
contribution of these components to the total spectrum must be
quantitatively
characterized. This is the goal of this Section.

\subsection{The soft X-ray spectrum}
\label{sect_soft}

The soft ({\it i.e.}, below $\simeq$2~keV) emission in Markarian~3 is
dominated by an AGN-photoionized plasma
\cite{bianchi05b,pounds05}, probably associated with the extended E-W structure
observed by the {\it Chandra}/ACIS \cite{sako00}. This component
contributes only a few percent to the flux in the energy range
used for spectral fitting in the following sections. However,
even this small contribution may significantly bias the best-fit
results if not properly taken into account.
We therefore decided to include
components describing the soft X-ray emission in all the models
described in this paper.

\begin{figure}
  \includegraphics[height=85mm,angle=-90]{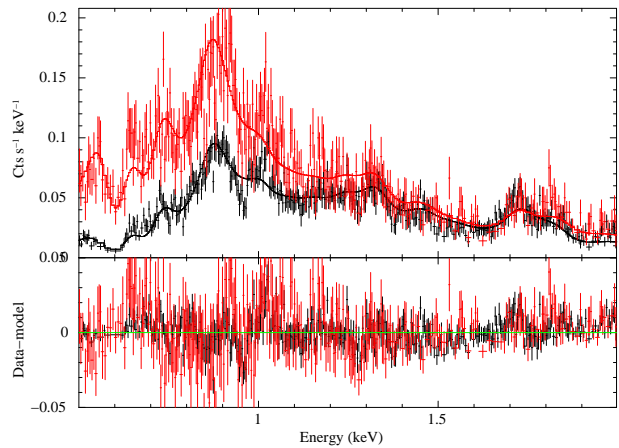}
  \caption{
  {\it Upper panel}; {\it Suzaku} BI ({\it red}) and FI-XIS ({\it black})
  Markarian~3 2005 spectra
  ({\it crosses}) and best-fit model ({\it line})
  in the 0.5--2~keV energy band. {\it Lower panel}: residuals against
  the best-fit. Spectral units in this and similar subsequent figures
  are counts per second per energy (keV).
  }
  \label{fig_soft2005suzaku}
\end{figure}

In order to achieve a model of the soft X-ray emission in Markarian~3,
we fit the {\it Suzaku} XIS spectra of the deep 2005 observation
in the 0.5--2~keV energy range (see Awaki et al. 2008 on the data reduction).
A good spectral fit is obtained
through the combination of one collisionally ionized optically thin plasma
({\tt apec}, Foster et al. 2012),
a power-law, and two photoionized components, seen
through a photoelectric absorption column density of
9.7$\times$10$^{20}$~cm$^{-2}$, equal to
the estimated contribution of neutral gas in our Galaxy \cite{kalberla05}.
We calculated the emission from a photoionized plasma using {\sc Xstar}
\cite{kallman14}, after
generating a level population file appropriate for the
Markarian~3 X-ray spectral energy distribution (energy index, $\alpha$,
-0.8), and using a
density $n_e$=10$^4$~cm$^{-3}$, an average value of the spatially resolved
model for the Narrow Line Region (NLR)
gas in Markarian~3 \cite{collins09}.
In this scenario, the power-law may be due to electron scattering, not
included in the {\sc Xstar} calculation of the reflection spectrum.
The 2005 XIS soft X-ray spectrum and
best-fit model are shown in Fig.~\ref{fig_soft2005suzaku}.
The best-fit parameters of the soft X-ray model
are shown in Tab.~\ref{tab_soft}.
\begin{table}
\caption{Best-fit parameters
of the soft X-ray model employed in this paper.}
\label{tab_soft}
\begin{tabular}{ll} \hline \hline
\multicolumn{2}{l}{Photoionized components} \\
$\log (\xi_{1,cgs})$ & $<$-0.7 \\
$N_{H,1}$ & 1.2$\times$10$^{21}$~cm$^{-2}$$^{\dag}$ \\
$\log (\xi_{2,cgs})$ & $2.00 \pm^{0.15}_{0.08}$ \\
$N_{H,2}$ & ($3.0 \pm 1.5$)$\times$10$^{22}$~cm$^{-2}$$^{\dag}$ \\
\multicolumn{2}{l}{Collisional component} \\
$kT$ & $0.83 \pm 0.02$~keV \\
\multicolumn{2}{l}{Power-law} \\
$\Gamma$ & $1.66 \pm 0.07$ \\
\multicolumn{2}{l}{Total observed flux} \\
$\log (F_{0.5-2 \ keV, cgs})$ & $-12.157 \pm 0.006$ \\ \hline \hline
\end{tabular} 

\noindent
$^{\dag}$unconstrained in the 10$^{20-24}$~cm$^{-2}$ range.
\end{table}
The fit is acceptable (C/$\nu$=1161.4/813). However,
an emission-line feature around
1.7--1.8~keV remain unaccounted for.
Formally, it can be fit with an unresolved Gaussian profile with
centroid energy $E$=1.744$\pm$0.007~keV and intensity
$I$=(6.5$\pm$1.0)$\times$10$^{-6}$~photons~cm$^{-2}$~s$^{-1}$.
It could be due either
to a 3p$^1$$\rightarrow$1s$^1$ transition of Mg{\sc XII}, or,
more likely, to known calibration uncertainties
at the energies of the XIS detector Silicon edge or escape peak.
The addition of further photoionized or collisionally ionized components,
even if we allow the elemental abundance to
be a free parameter in the fit,
yield a negligible further improvement in the quality of the fit.
Given the uncertainties on its origin, we do not discuss
this feature in this paper, and do not include it in the
soft X-ray model. 

We then applied this soft X-ray baseline model to the
0.5--2~keV X-ray XIS
spectra taken during the 2014-2015 observational campaign.
The fits are satisfactory,
just allowing for an overall normalization constant, $C_{soft}$, to vary.
The best for values of $C_{soft}$ are $0.99 \pm 0.03$,
$0.96 \pm 0.03$, and $0.78 \pm 0.03$, for Obs.\#709022010 to 709022040,
respectively. Leaving the
power-law spectral index free yields a significantly
different value from that measured in 2005 only
for Obs.\#709022030: $\Gamma_{030}$=1.97$\pm^{0.26}_{0.15}$,
but the difference in the quality of the fit is marginal ($\Delta
\chi^2 = 0.6$). This suggests only
a small contribution of partial covering to the
soft X-ray emission.
Varying either the warm emitter
ionization parameter or column density, or the power-law
normalization at each epoch yields values always consistent with
the common best-fit.

The power-law component contributes over two orders-of-magnitude
more than the other components to the flux
above 4~keV. Consequently, in the model described in the following
we kept
the soft X-ray baseline model parameters frozen to the values determined from
the 0.5--2~keV {\it Suzaku}/XIS fit whenever {\it Suzaku} spectra are
used; and allowed the normalization of the soft power-law to vary
within a dynamical range of $\pm$20\% whenever {\it NuSTAR} data are
analyzed alone, or together with the EPIC-pn spectra. The EPIC-pn
exposures discussed in this paper are too short to constrain
the properties of the soft X-ray emission.

A detailed discussion of the variability of the soft X-ray spectrum
of Markarian~3 is deferred to a future paper. More details on the
soft X-ray spectroscopy can be found in: Sako et al. (2000),
Bianchi et al. (2004), Awaki et al. (2008), and Yaqoob et al. (2015).

\subsection{{\tt mytorus}-based models}
\label{sect_mytorus}

We refrain in this paper from using purely empirical models to describe the
broad-band X-ray spectra obtained during the 2014--2015 campaigns. The
unprecedented combination of broad-band X-ray coverage and high-sensitivity
above 10~keV, coupled with the possibility of constraining the observables
of the Fe K$_{\alpha}$ lines in those observations where {\it NuSTAR}
and a CCD instrument
were used quasi-simultaneously, allow us to constrain the parameter space
of geometrically motivated models describing optically thick reprocessing of
the primary continuum in this heavily obscured AGN. In this Section we
make use of {\tt mytorus} \cite{murphy09}, describing Compton-thick
reprocessing in a toroidal geometry with a fixed half-opening angle
$\theta_{op}$=60$^{\circ}$ and solar abundances.
The assumed opening angle in {\tt mytorus}
is in good agreement with that expected if the torus
in Markarian~3 follows the anti-correlation between X-ray luminosity and
covering fraction observed in bright nearby Seyferts \cite{brightman15}.
{\it NuSTAR} 
successfully constrained the properties of the AGN ``torus'' in a number
of heavily obscured AGN
\cite{arevalo14,balokovic14,puccetti14,gandhi14,bauer15,brightman15,rivers15,marinucci16}.

Following Yaqoob et al. (2015), we fit the data using a ``decoupled
scenario'', described by the following equation:
$$
M(E)=S(E) + A(N_{H,abs},E) \times (N_{po} E^{\Gamma}) + \nonumber
$$
\begin{equation}
\label{eq_mytorus}
+ C_{sc} \times [R_c(N_{H,sc}) + R_l(N_{H,sc})] + \Sigma_i G_i(E_l)
\end{equation}
where $S(E)$ is the soft X-ray model described in Sect.~\ref{sect_soft};
 A($N_{H,abs}, E$; model {\tt mytorus\_Ezero\_v00.fits})
is an energy-dependent multiplicative factor describing the
absorption of the primary continuum along the line-of-sight through a
column density $N_{H,abs}$, including the effect of Compton scattering.
We parameterized
the primary continuum with a simple power-law
with photon index $\Gamma$;
$R_c$ and $R_l$ (models
{\tt mytorus\_scatteredH500\_v00.fits} and
{\tt mytl\_V000010nEp000H500\_v00.fits}) represent the scattered
emission (in the continuum and in the line) due to the global column
density $N_{H,sc}$; $C_{sc}$ takes into account
differences between the transmitted primary continuum level
and that illuminating the reflecting clouds; and $G_i$ are Gaussian profiles
describing a Fe{\sc xxvi}
resonant
absorption line, and a Fe{\sc xxv} emission line (see Sect.~\ref{sect_hetg}). 
In this scenario
the inclination angle of the Compton scattered components is set to
0$^{\circ}$ \cite{yaqoob12}.
The inclination angle of the absorption table is a dummy
parameter (fixed to 90$^{\circ}$ in the fit).
   
In the fitting procedure, we imposed that parameters not expected to vary
over the time span covered by the monitoring campaign
are tied together at all epochs (``global parameters'', hereafter). They are:
$C_{s,1}$ (the subscript ``1'' indicates the first epoch in our
monitoring campaign), $N_{H,sc}$.
The centroid energy of the absorption line was also frozen to 6.96~keV
in the fit. The following astrophysical
parameters were allowed to vary independently in each epoch: $\Gamma$,
the power-law normalization $N_{po}$,
$N_{H.abs}$ and the intensity of the absorption line
$I_{abs}$ (``epoch-dependent'' parameters hereafter).
Finally, we have made the assumption that the flux of the Compton
scattering and the intensity of the fluorescent line do not vary during
our monitoring campaign, after verifying that this assumption is consistent
with the data.
We will refer to this model as the ``{\tt mytorus} baseline
model'' hereafter.
 
Fitting simultaneously the 26 spectra of the 2014--2015 Markarian~3 campaign
with this model yields a good fit, with C/$\nu$=32004.0/35972.
The residuals are featureless (Fig.~\ref{residuals_absorption}
and \ref{residuals_absorption_2}) in all observations.
\begin{figure*}
  \includegraphics[height=75mm,angle=-90]{f_20140907.ps}
  \includegraphics[height=75mm,angle=-90]{f_20140914.ps}
  \includegraphics[height=75mm,angle=-90]{f_20141001.ps}
  \includegraphics[height=75mm,angle=-90]{f_20141009.ps}
  \includegraphics[height=75mm,angle=-90]{f_20141023.ps}
  \includegraphics[height=75mm,angle=-90]{f_20150319.ps}
  \includegraphics[height=75mm,angle=-90]{f_20150322.ps}
  \includegraphics[height=75mm,angle=-90]{f_20150405.ps}
  \caption{Spectra ({\it upper panels}) and residuals in units of data
    versus model ratio ({\it lower panels}) when the {\tt mytorus} baseline
    model is fit
    to the epochs
    between 2014 September 7 and 2015 March 5. Where two spectra are shown,
    they correspond to the NuSTAR/FPM units; where three spectra are shown,
    the EPIC-pn spectrum is included; where four spectra are shown,
    the XIS spectra (BI and FI) are included.
    }
  \label{residuals_absorption}
\end{figure*}
\setcounter{figure}{5}
\begin{figure}
  \includegraphics[height=85mm,angle=-90]{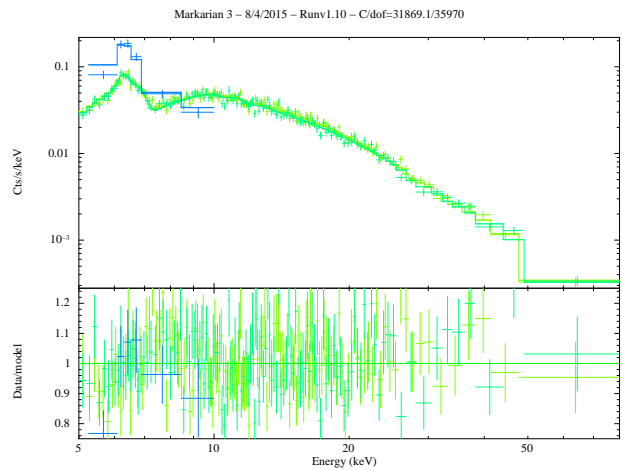}
  \caption{continued.
    }
  \label{residuals_absorption_2}
\end{figure}
Possible enhancements do not significantly further improve
the quality of the fit: using
{\tt mytorus}
components corresponding to lower values of the primary continuum high-energy
cut-off (500~keV in the baseline model; this is consistent with the
lower limit on the cut-off energy that we can set with out data
at the 99\% confidence level for two interesting
parameters);
allowing the centroid energy of the absorption line to be free
(the best-fit is always consistent with the rest-frame energy of
Fe{\sc xxvi}); or
using a ``coupled scenario'', {\it i.e.} a scenario where the column
density and the inclination angles associated to all {\tt mytorus}
model components are tied together (the hypothesis of a fully
homogeneous torus). At variance with the results of Yaqoob et al.
(2015), the quality of the fit in ``coupled mode''
is comparable to that obtained in
``decoupled mode'' (C/$\nu$=31958.3/35973), there is no systematic
curvature in the residuals, and the best-fit values
of the intrinsic power-law photon index (1.77-1.93) are not particularly
flat. However, it remains true that the best-fit values of the column
density vary among different epochs (in the range
2.64--3.26$\times 10^{24}$~cm$^{-2}$,
with a typical statistical error $\simeq$0.03$\times 10^{24}$~cm$^{-2}$).
Therefore, we rule out this scenario on astrophysical grounds, because it is
difficult to explain how the {\it global} column density of the
Compton scattering material could vary on time scales as short as a few days.

Table~\ref{table_mytorusbaseline} reports the epoch-dependent parameters
in the {\tt mytorus} baseline scenario.
The line-of-sight column density
varies between 7.5 and 9$\times$10$^{23}$~cm$^{-2}$,
in agreement with the re-classification of Markarian~3 as ``Compton-thin''
proposed by Yaqoob et al. (2015); the primary
spectral index between 1.65 and 1.85, the normalization fluctuates
about 30\% around its average.
\begin{table*}
\caption{Best-fit parameters for the epoch-dependent best-fit parameters discussed in this paper.}
\label{table_mytorusbaseline}
\begin{tabular}{lccccccccc} \hline \hline
epoch & 1 & 2 & 3 & 4 & 5 & 6 & 7 & 8 & 9 \\
(YY/MN/DD) & 14/09/07 & 14/9/14 & 14/10/01 & 14/10/09 & 14/10/23 & 15/03/19 & 15/03/23 & 15/04/05 & 15/04/08 \\ \hline
\multicolumn{10}{l}{{\tt mytorus}} \\
$\Gamma$ & 1.66$\pm^{0.03}_{0.01}$ & 1.69$\pm^{0.02}_{0.01}$ & 1.73$\pm^{0.02}_{0.01}$ & 1.72$\pm^{0.03}_{0.01}$ & 1.65$\pm^{0.02}_{0.01}$ & 1.81$\pm^{0.07}_{0.07}$ & 1.78$\pm^{0.06}_{0.06}$ & 1.85$\pm^{0.04}_{0.03}$ & 1.81$\pm^{0.03}_{0.01}$ \\
$N$$_{po}^a$ & 2.98$\pm^{0.22}_{0.22}$ & 3.67$\pm^{0.24}_{0.04}$ & 2.95$\pm^{0.09}_{0.23}$ & 3.58$\pm^{0.25}_{0.21}$ & 2.89$\pm^{0.26}_{0.26}$ & 3.37$\pm^{0.21}_{0.56}$ & 3.35$\pm^{0.18}_{0.18}$ & 4.46$\pm^{0.31}_{0.41}$ & 4.08$\pm^{0.23}_{0.18}$ \\
$N$$_{H,abs}^b$ & 0.77$\pm^{0.01}_{0.01}$ & 0.77$\pm^{0.01}_{0.01}$ & 0.81$\pm^{0.01}_{0.01}$ & 0.75$\pm^{0.01}_{0.01}$ & 0.74$\pm^{0.01}_{0.01}$ & 0.89$\pm^{0.01}_{0.02}$ & 0.87$\pm^{0.01}_{0.01}$ & 0.94$\pm^{0.01}_{0.02}$ & 0.90$\pm^{0.01}_{0.02}$ \\ \hline
\multicolumn{10}{l}{Ikeda} \\
$\Gamma$ & 1.76$\pm^{0.02}_{0.01}$ & 1.80$\pm^{0.04}_{0.02}$ & 1.81$\pm^{0.01}_{0.03}$ & 1.80$\pm^{0.02}_{0.02}$ & 1.73$\pm^{0.02}_{0.02}$ & 1.88$\pm^{0.04}_{0.00}$ & 1.85$\pm^{0.01}_{0.01}$ & 1.89$\pm^{0.01}_{0.02}$ & 1.87$\pm^{0.01}_{0.02}$ \\
$N$$_{po}^a$ & 2.79$\pm^{0.10}_{0.08}$ & 3.81$\pm^{0.46}_{0.15}$ & 2.53$\pm^{0.13}_{0.21}$ & 3.22$\pm^{0.21}_{0.22}$ & 2.70$\pm^{0.13}_{0.14}$ & 2.93$\pm^{0.31}_{0.29}$ & 2.74$\pm^{0.15}_{0.15}$ & 3.47$\pm^{0.24}_{0.43}$ & 3.44$\pm^{0.18}_{0.18}$ \\
$N$$_{H,abs}^b$ & 0.86$\pm^{0.01}_{0.01}$ & 0.92$\pm^{0.01}_{0.01}$ & 0.88$\pm^{0.01}_{0.10}$ & 0.86$\pm^{0.01}_{0.01}$ & 0.84$\pm^{0.01}_{0.01}$ & 1.01$\pm^{0.02}_{0.03}$ & 0.95$\pm^{0.02}_{0.02}$ & 1.08$\pm^{0.02}_{0.02}$ & 1.03$\pm^{0.01}_{0.03}$ \\
$I$$_{abs}^c$ & -2.0$\pm^{ 0.5}_{ 0.5}$ & -2.0$\pm^{ 0.5}_{ 0.5}$ & -0.7$\pm^{ 0.3}_{ 0.3}$ & -2.1$\pm^{ 0.5}_{ 0.5}$ & $>-0.5$ & -0.2$\pm^{ 0.2}_{ 0.5}$ & -0.4$\pm^{ 0.4}_{ 0.4}$ & $>-0.5$ & $>-0.5$ \\
$EW$$^d$ & -20$\pm^{  5}_{  6}$ & -15$\pm^{  4}_{  4}$ &  -8$\pm^{  4}_{  5}$ & -18$\pm^{  5}_{  6}$ & $>-6$ &  -1$\pm^{  1}_{  4}$ &  -4$\pm^{  4}_{  5}$ & $>-6$ &   $>-6$ \\
$L$$_{AGN}$$^e$ & 4.21$\pm^{0.15}_{0.09}$ & 5.44$\pm^{0.65}_{0.55}$ & 3.52$\pm^{0.13}_{0.25}$ & 4.62$\pm^{0.22}_{0.30}$ & 4.25$\pm^{0.20}_{0.21}$ & 3.71$\pm^{0.28}_{0.57}$ & 3.62$\pm^{0.14}_{0.47}$ & 4.32$\pm^{0.21}_{0.21}$ & 4.41$\pm^{0.15}_{0.36}$ \\ \hline \hline
\end{tabular}
 
\noindent
$^a$units of 10$^{-2}$~photons~cm$^{-2}$~s$^{-1}$~keV$^{-1}$ at 1~keV
 
\noindent
$^b$units of 10$^{24}$~cm$^{-2}$
 
\noindent
$^c$units of 10$^{-5}$~photons~cm$^{-2}$~s$^{-1}$
 
\noindent
$^d$units of eV
 
\noindent
$^e$2--10~keV luminosity of the transmitted component only, corrected for absorption, in units of 10$^{43}$~erg~s$^{-1}$
\end{table*}
The {\tt mytorus}
baseline model requires a global column density
$N_{H,sc}$=($1.16 \pm^{0.08}_{0.11}$)$\times$10$^{23}$~cm$^{-2}$. This
is
a factor of $\simeq$7 lower than the line-of-sight column density.
A similar torus structure was inferred recently for ESO138-G0001
\cite{decicco15}.
Any
deviation from a constant, isotropic illumination of the scattering
clouds is, at most, modest: $C_{sc,1}$=($1.15 \pm 0.08$).
The intensity of the Fe{\sc xxvi} absorption line
is a factor of two larger in 2014 than in 2015.
Its Equivalent Width (EW) --
calculated only against the primary transmitted continuum -- is 10--20~eV
in 2014, decreasing to $\approxlt$6~eV in 2015.

\subsection{Ikeda-based models}
\label{sect_ikeda}

Ikeda et al. (2009; I09 hereafter) developed a Monte
Carlo model to study the reflection-dominated spectra of
heavily obscured AGN.
I09 assume a spherical-toroidal geometry with a varying opening angle as opposed to
the purely toroidal geometry of {\tt mytorus}.
The models therefore differ in their detailed predictions \cite{ricci14}.
This sub-section discusses the application of the I09 models to
the 2014--2015 Markarian~3 monitoring campaign, and compares
its results with those obtained with
{\tt mytorus} in Sect.~\ref{sect_mytorus}.

We run a series of fits in a configuration
equivalent to the {\tt mytorus} ``decoupled
mode'':
$$
M(E)=S(E) + A(N_{H,abs},E) \times (N_{po} E^{\Gamma}) + \nonumber
$$
$$
+ C_{sc,1} \times [R_1(N_{H,sc},\theta_{incl},\theta_{op}) + R_2(N_{H,sc},\theta_{incl},\theta_{op}) +
$$
\begin{equation}
\label{eq_ikeda}
+ R_{Fe}(N_{H,sc},\theta_{incl},\theta_{op})] + \Sigma_i G_i(E_l)
\end{equation}
where $R_2$ indicates the light reflected from the inner far side
of the torus, and $R_1$ the remaining contribution to the continuum
Compton-scattered emission. The Fe fluorescence component is embedded in
$R_{Fe}$.
The free parameters
in the Ikeda-based model are the same as in
{\tt mytorus}, with the addition of the torus
half-opening angle ($\theta_{op}$) and inclination angle ($\theta_{incl}$).
Moreover, the I09 line emission model does not include the Fe K$_{\beta}$ or the
Ni emission lines. We thus included them in the model through phenomenological Gaussian profiles.
The best-fit ratio between Fe K$_{\beta}$ 
and the Fe K$_{\alpha}$ obtained with the models discussed in this Section
is $\simeq$7\%, with a statistical error of about 50\%.
This value is lower, but still
in broad agreement with the expectations from
atomic physics \cite{molendi03}. The constant ratio
between the Ni K$_{\alpha}$ and the Fe K$_{\alpha}$ ($\simeq$20\% with a 50\% error bar)
is instead larger than expected from the
nickel-to-iron cosmic abundance (0.055-0.075). We
refer to this model as the ``baseline Ikeda model''
hereafter.

The fit to the 26 spectra of our monitoring campaign
is marginally better than with {\tt mytorus}-based models
(C/$\nu$=31922/35969).
In terms of best-fit parameter values, the main differences with respect to the
{\tt mytorus} baseline can be summarized as follows: 

\begin{itemize}

\item a 15\% higher
  line-of-sight column density (cf. Table~\ref{table_mytorusbaseline}),
  with a similar time variability pattern, but still consistent with
  a Compton-thin absorber;

\item a
  global column density larger by a factor of about 2 than the
  transmitted column density
  $N_{H,sc}$=($2.01 \pm^{0.11}_{0.08}$)$\times$10$^{24}$~cm$^{-2}$,
with a scattering fraction $C_{cs,1}$=1.00$\pm^{0.05}_{0.08}$; and

\item a relatively wide torus ($\theta_{op}$=66.0$\pm$0.4$^{\circ}$) seen at
  an almost grazing angle ($\theta_{incl}$=70$\pm$3$^{\circ}$)

\end{itemize}

The best-fit parameters of the Fe{\sc xxvi} absorption line are very
similar to those obtained with the {\tt mytorus} baseline model.
In Table~\ref{table_mytorusbaseline} we show the line intensity $I_{abs}$
and EW (calculated against the transmitted continuum only)
obtained with the Ikeda baseline model.

\subsection{{\tt torus}-based models}
\label{sect_torus}

\begin{figure}
  \hspace{-1.0cm}
  \hbox{
    \includegraphics[height=95mm, angle=90]{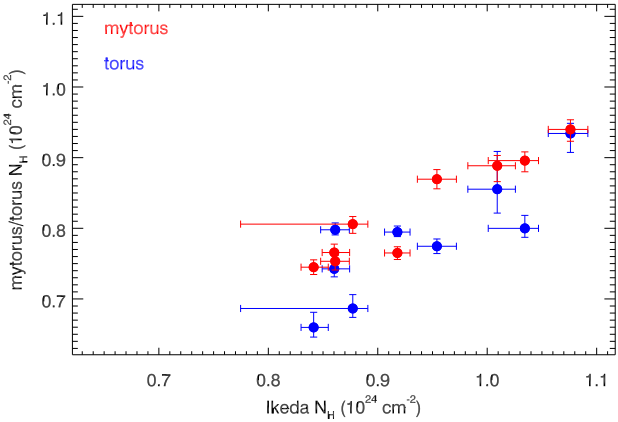}
  }
  \caption{Comparison between the column density in transmission measured at
    the nine epochs of the 2014--2015 Markarian~3 monitoring campaign using I09
    (X-axis), {\tt mytorus} (y-axis, {\it red points}), and {\tt torus}
    (y-axis, {\it blue points}).
    }
  \label{fig_varpattern}
\end{figure}

We also fit the spectra of our monitoring campaign with the
{\tt torus} model \cite{brightman11}, a Monte Carlo simulation code
based on the method of George \& Fabian (1991) that assumes a
biconical toroidal distribution similar to that employed in I09.
This model can be used only in ``coupled mode'', {\it i.e.} linking
column density in transmission and in reflection.
As already discussed in Sect.~\ref{sect_mytorus},
{\tt torus} cannot therefore be applied
to our data in a fully self-consistent way.
In spite of that,
the column density best-fit values are in the same range as
those derived employing {\tt mytorus}, and follow a very similar variability
pattern as seen using the {\tt mytorus} and the Ikeda models
(Fig.~\ref{fig_varpattern}). 
The torus properties derive with {\tt torus} are in good agreement with
that derived with I09. For instance,
the best-fit opening angle
is $\theta_{op}$=58$\pm$3$^{\circ}$.
This value is in good agreement with the correlation
between X-ray luminosity and torus covering fraction measured
in a sample of nearby heavily obscured AGN by {\it NuSTAR} \cite{brightman15}.
The corresponding inclination angle ($\theta_{incl} \ge$83$^{\circ}$) is,
however, significantly larger than with the I09 model, disfavoring
a grazing view.

\subsection{Comparison with previous observations}

\subsubsection{Suzaku 2005}

Table~\ref{tab2005} lists the best-fit parameters obtained when the
\begin{table}
\caption{Best-fit parameters when the {\tt mytorus} baseline model
is applied to the September 2005 {\it Suzaku} observation of Markarian~3.}
\label{tab2005}
\begin{tabular}{lc} \hline \hline
$\Gamma$ & $1.74 \pm^{0.008}_{0.021}$ \\
$N_{po}$ (10$^{-2}$~ph~cm$^{-2}$~s$^{-1}$~keV$^{-1}$) & $1.58 \pm^{0.02}_{0.03}$ \\
$N_{H,abs}$ (10$^{24}$~cm$^{-2}$) & $1.24\pm0.03$ \\
$I_{abs}$ (10$^{-5}$~ph~cm$^{-2}$~s$^{-1}$) & $>-0.3$ \\
$C_{sc}$ & $1.24\pm^{0.16}_{0.05}$ \\ \hline \hline
\end{tabular}
\end{table}
baseline Ikeda model is applied to the deep {\it Suzaku} observation
of Markarian~3 in September 2005 (see I09
for a discussion on the application of the Ikeda model to the same
observation).
In performing the fit, we constrained the torus structural parameters
($N_{H,sc}$, $\theta_{op}$, $\theta_{inc}$) within the confidence intervals
determined from the 2014-2015 monitoring campaign, because they were
determined with better precision than in previous studies.
This assumes that
the global structure of the pc-scale
torus has not changed over the last decade. The fit is good
(C/$\nu$=2952/2862).
The continuum parameters
are in broad agreement with those obtained during the
2014--2015 campaign. The line-of-sight column density is larger by a factor
$\simeq$70\%, changing the
nominal source classification to a borderline Compton-thick object.
The power-law
normalization is lower by a factor $\simeq$2, without this change
being reflected in a proportional increase of the scattering
fraction. No absorption line is
detected in the energy range consistent with resonant transitions of
He- or H-like iron.

\subsubsection{{\it Chandra}/HETG}
\label{sect_hetg}

We downloaded from the TGCAT archive
\cite{huenemoerder11} the High-Energy
Transmission Grating
(HETG)
spectra of Markarian~3 accumulated during the course of the {\it Chandra} mission,
together with their associated background spectra and responses.
These
correspond to nine exposures\footnote{Observations numbers: 873, 12874, 12875, 13254, 13261,
13263, 13264, 13406, and 14331} for a total net exposure time of about 778600 seconds.
They span a time range between March 2000 and April 2011. 
The Medium Energy Grating (MEG) is insensitive above 5 keV. We therefore merged together
the High Energy Grating (HEG) first order (positive and negative) spectra using the
{\tt addascaspectra} tool. The count rate in the 4--8~keV energy range is $(4.40 \pm 0.07) \times 10^{-3}$~s$^{-1}$,
corresponding to 3402 net total counts. We determined
that no variation in the continuum level below 6 keV, or in the intensity of the Fe
K$_{\alpha}$ line is apparent in the spectra extracted from the individual {\it Chandra}
observations.

\begin{figure}
  \includegraphics[height=85mm, angle=-90]{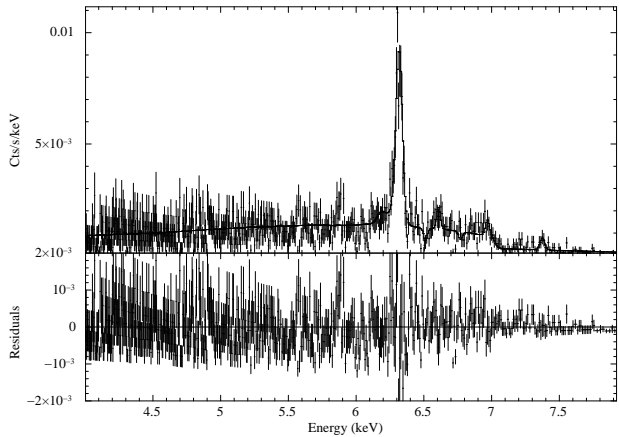}
  \caption{
    Markarian~3 {\it Chandra}/HETG time-averaged spectrum ({\it crosses, upper panel})
    and residuals against the best-fit model ({\it lower panel}). The {\it solid line} in the {\it upper panel} 
    represents the Ikeda model based
    best-fit (details in text).
    }
  \label{fig_hetg}
\end{figure}

The combined, time-averaged spectrum was fit in the 4--8 keV band using the
Ikeda baseline continuum,
and adding emission and absorption lines (modeled with simple Gaussian profiles) if statistically
motivated at a confidence level larger than 90\% for one interesting parameter.
We note that the I09 model includes the Compton shoulder. There is therefore no
need of fitting this feature with {\tt ad hoc} phenomenological profiles.
In order not to over-fit the data,
we fixed the torus structural parameters
($N_{H,sc}$, $\theta_{op}$, $\theta_{inc}$) to the values
determined with the data of the 2014-2015 monitoring campaign.
The final
model requires one unresolved emission line, and two
unresolved absorption lines. Their energies are consistent with Fe{\sc xxv}
recombination (6.7~keV, unconstrained), and Fe{\sc XXV} ($6.60 \pm^{0.06}_{0.08}$~keV)
and Fe{\sc XXVI} (6.96~keV, unconstrained) resonant absorption,
respectively.
The best-fit is acceptable ($C/\nu$=672.4/584), and does not show any systematic
residual features (Fig.~\ref{fig_hetg}),
especially in the iron band. The high-resolution allows us also to
constrain the width of the Fe K$_{\alpha}$ line. We did this by applying
an energy-independent
Gaussian convolution kernel to the best-fit model.
Its best-fit width is $\sigma = 15\pm4$~eV, corresponding to
$700\pm200$~km~s$^{-1}$ in velocity space.
Table~\ref{tab_hetg} shows
\begin{table}
  \caption{Best-fit parameters when the
    model shown in Fig.~\ref{fig_hetg}
    is applied to the {\it Chandra}/HETG time-averaged spectrum of Markarian~3.
  }
\label{tab_hetg}
\begin{tabular}{lc} \hline \hline
$\Gamma$ & $>2.46$ \\
$N_{po}$ (10$^{-2}$~ph~cm$^{-2}$~s$^{-1}$~keV$^{-1}$) & $8 \pm 3$ \\
  $N_{H,abs}$ (10$^{24}$~cm$^{-2}$) & $1.29\pm^{0.18}_{0.03}$ \\
$C_{sc}$ & $2.17 \pm^{0.13}_{0.28}$ \\
  $I_{emi}$ (10$^{-5}$~ph~cm$^{-2}$~s$^{-1}$) & $0.4 \pm 0.2$ \\
  $I_{abs,FeXXV}$ (10$^{-5}$~ph~cm$^{-2}$~s$^{-1}$) & $-0.22\pm^{0.11}_{0.12}$ \\
  $I_{abs,FeXXVI}$ (10$^{-5}$~ph~cm$^{-2}$~s$^{-1}$) & $>-0.16$ \\
  $\sigma$ (eV) & $15\pm 4$ \\
  \hline \hline
\end{tabular}
\end{table}
the best-fit parameters.

\section{Discussion}
\label{sect_discussion}

\subsection{The origin of the hard X-ray variability in Markarian~3}
\label{sect_originvariability}

\begin{figure}
  \includegraphics[height=85mm, angle=90]{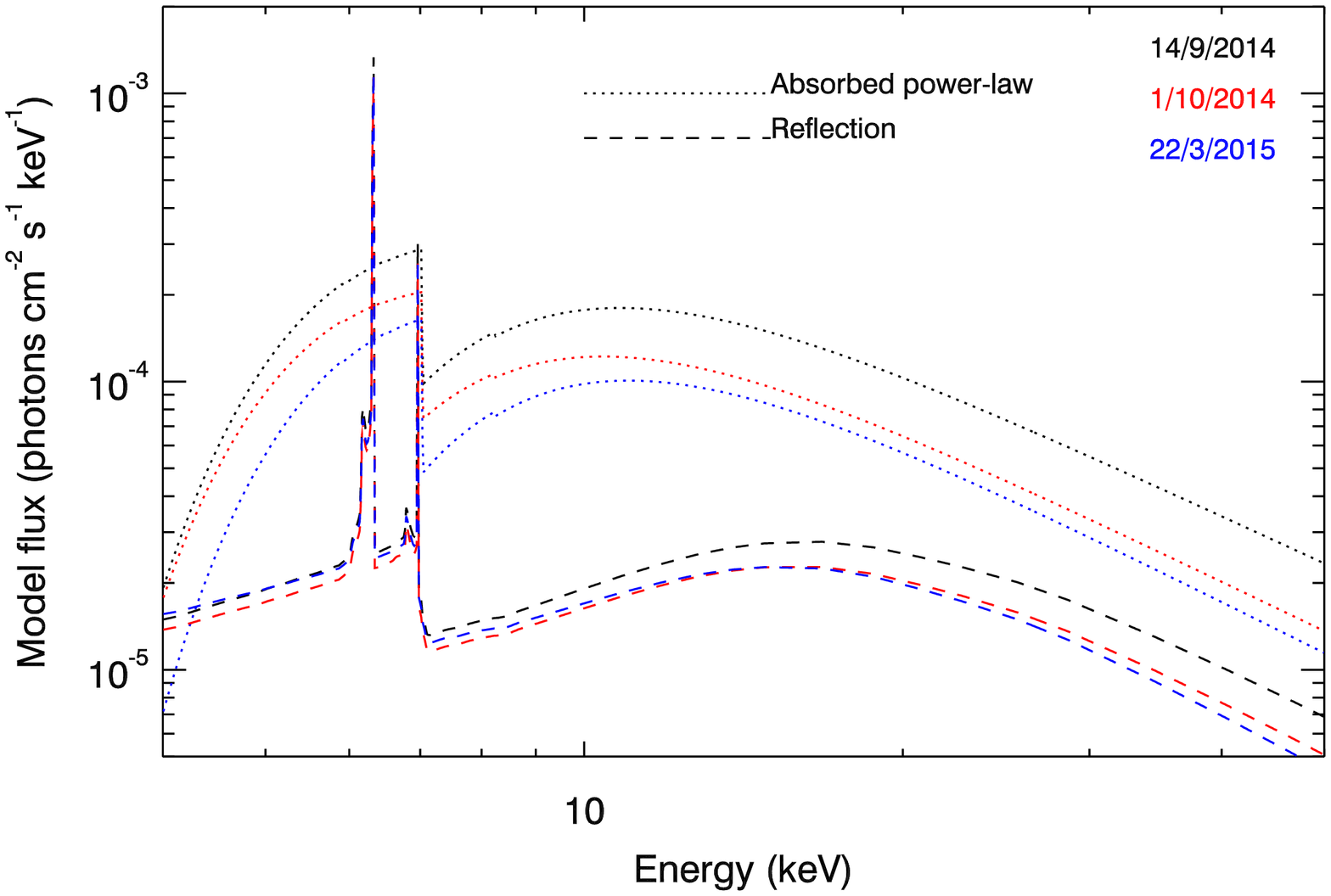}
  \includegraphics[height=85mm, angle=90]{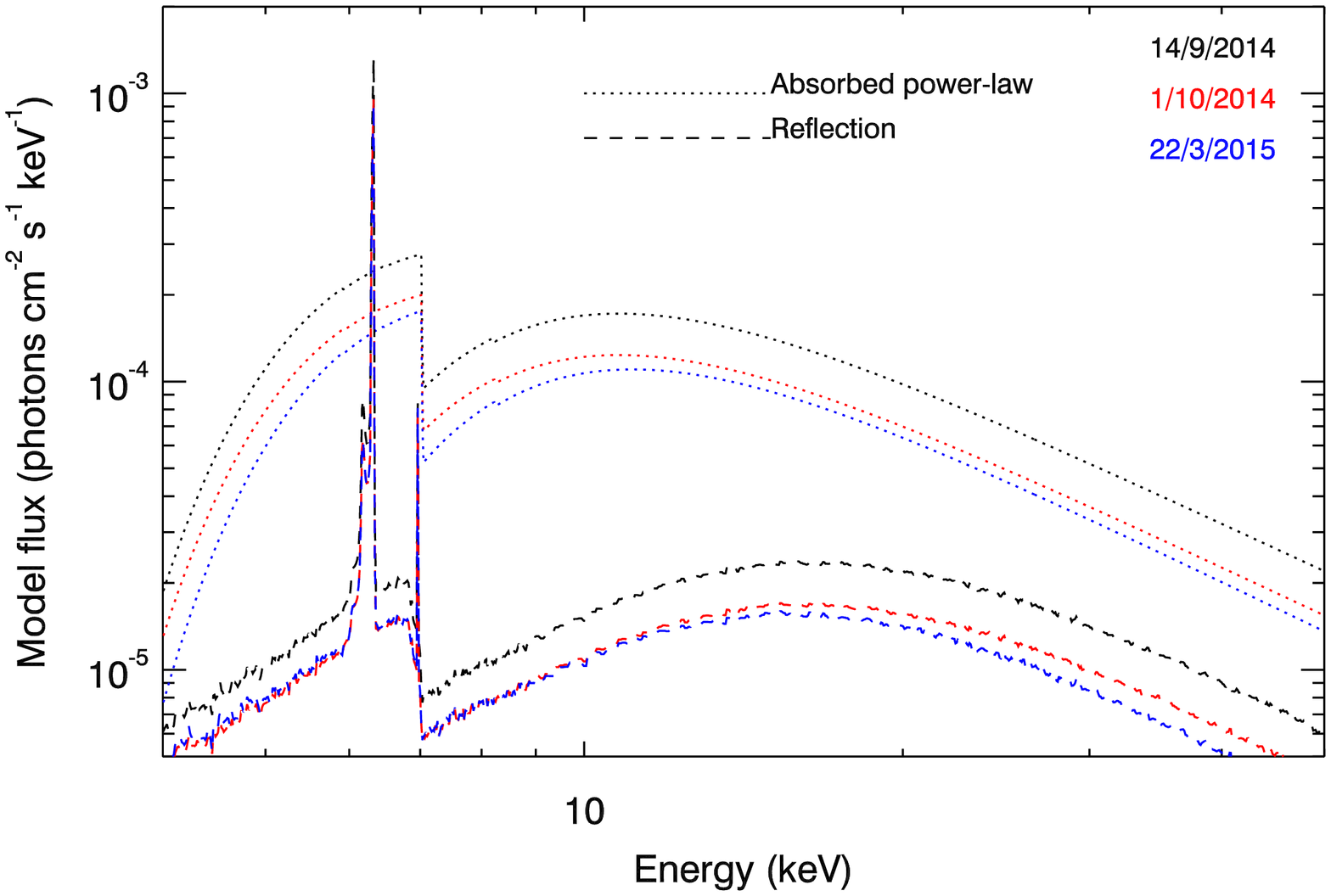}
  \caption{Best-fit models in three epochs for the {\it mytorus}
    ({\it top}) and the Ikeda ({\it bottom}) best-fit models
    in three epochs: 2014 September 14 ({\it black}), October 1 ({\it red}), and 2015 March 22
    ({\it blue}). The {\it dotted line} represents the transmitted component, the {\it dashed line}
    the reflected component.
    }
  \label{fig_models}
\end{figure}

In our previous paper on the recent X-ray history of 
Markarian~3, we (G12) proposed that the variability pattern of the
4--5~keV flux could be used to put constraints on the nature of the
reprocessing matter.
The results of the monitoring campaign reported in this paper
rule out this claim, and contradict the idea that the spectral variability
in this energy band can be used to constrain the properties
of a Compton-thick reflector. Thanks to the accuracy in the
spectral deconvolution allowed by the {\it NuSTAR} unprecedented combination of energy
bandpass and spectroscopic quality above 10~keV, and
to the usage of self-consistent physical models of the reprocessing,
we achieve a more robust interpretation of the
variability pattern. 
The
continuum variability in the 4--50~keV band can be explained
as due to the variation of the primary
emission, (both in spectral index $\Gamma$ and normalization
$N_{po}$) (Fig.~\ref{fig_models}).
As we will show in Sect.~\ref{sect_2014} and \ref{sect_2015},
variability of the intervening absorber column density also
contributes to the observed continuum variability.
In these models, only the iron line traces the
reflection component, while the 4--5~keV flux traces primarily
that of the primary continuum in transmission with some dilution from
the underlying reflection component. G12 had already remarked that the
intensity of the Fe K$_{\alpha}$ lines does not track the 4--5~keV
flux as would have been expected if they were both produced in the
same reprocessing material. Fig.~\ref{fig_models} provides us with a
simple explanation of this observational fact.

Our results are in agreement with the systematic study of the
X-ray variability in a sample of Seyfert~2 discussed
by Hern\'andez-Garc\'ia et al. (2015). These authors interpret the
historical spectral variability in the XMM-Newton observations of
Markarian~3 as primarily due to variations of the normalization of
the primary power-law. The limited bandpass of the
XMM-Newton scientific payload
prevented them from constraining the additional contribution due
to variability
of the intervening absorber column density. This can be fully unveiled
in our campaign
thanks to the extended coverage and hard X-ray unprecedented sensitivity
of NuSTAR. In Hernandez-Garcia et al. (2015), Markarian~3 stands out as
the only AGN classified as Compton-thick through indirect indicators
showing variability in the 0.5--10~keV band. This lends additional
support to the classification of Markarian~3 as a Compton-thin object
(see also the discussion at the end of Sect.~\ref{sect_timeaveraged}).

\begin{figure}
  \includegraphics[height=65.5mm]{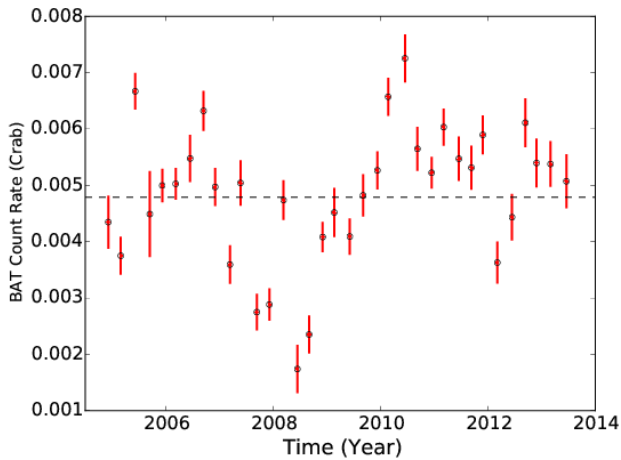}
  \caption{{\it Swift}/BAT light curve of Markarian~3 in the 14-195 energy band.
    }
  \label{batlc}
\end{figure}

As the {\it Swift}/BAT light curve shows (Fig.~\ref{batlc}), the
AGN in Markarian~3 in one of the most variable in the X-ray band
(G12). The overall dynamical range in the last decade has been
a factor of about eight, with a doubling time on the order of
1 year.

No significant intra-observation variability has been detected in
the high-energy {\it NuSTAR} light curves of Markarian~3,
at variance with, {\it e.g.},
NGC~4945 \cite{puccetti14}.
This is partly due to the lower statistics of the Markarian~3 data.
The 30--79~keV Markarian~3 count rates
during the 2014--2015 campaign were a factor of 2 to 10 lower
than in NGC~4945. More fundamentally, Markarian~3 has been estimated to host
a very massive black hole
($M_{BH}$$\sim$4.5$\times$10$^8$M$_{\odot}$; Woo \& Urry 2002). If 
its X-ray primary continuum follows the relation between
power-spectrum break time-scale $T_B$, black hole mass
and luminosity established for unobscured AGN \cite{mchardy06}:
$$
\log(T_B) = 2.1 \log(M_{BH,6}) - 0.98 \log(L_{bol,44}) - 2.28
$$
where $T_B$ is in days, $M_{BH,6}$ is in units of 10$^{6}$M$_{\odot}$ and
$L_{b,44}$ is in units of 10$^{44}$~erg~s$^{-1}$, $T_B$ is of the order
of $\simeq$170~days. This indicates that little power is expected
at frequencies comparable with the typical duration of {\it NuSTAR}
observations with respect to the long-term behavior.

We note in passing that a
slight improvement in the quality of the fit in the {\tt mytorus}
scenario is obtained if $C_{sc}$ is allowed to independently
vary for each epoch ($\Delta C$/$\Delta \nu$=46.9/8) with $C_{sc}$ varying
between 0.38$\pm$0.08 in the
highest flux state (2014 September 14) to 0.80$\pm$0.11, 0.73$\pm^{0.18}_{0.15}$,
and 0.77$\pm$0.12
in the lowest flux states (2014 October 1 and
2015 March 19 and 23).
However, it is hard to find a convincing astrophysical explanation for this finding.
The mathematical interpretation of $C_{sc}$ is
complex. One should
refrain from interpreting it as a simple covering fraction of the
optically thick reprocessing matter \cite{murphy09,yaqoob12}. It is rather a combination of the
variability and anisotropy properties of the illuminating continuum,
of the light crossing time delays between the illuminating source and
the reprocessing matter,
and of the detailed (and unknown) geometry of the reprocessing clouds.
{\it If} the change in the $C_{sc}$ values
were solely due to a change in the property of the
global reprocessing matter surrounding the central engine,
one needs to invoke a mechanism to, for instance, more than double
its column density or covering fraction
between the second and the third observations in 2014, separated by 17 days.
If due, for instance, to the onset of a disk
wind at the virial escape velocity, $v_e = (G M_{BH}/r)^{0.5}$,
and assuming that the wind clouds have to travel
a distance at least comparable to the distance from the illuminating source to
significantly contribute to the X-ray reflection,
this wind should be launched at
$r \sim$4$\times$10$^{15}$~cm, or 70 gravitational radii, $r_g$.
Disk winds launched in the innermost regions of the accretion flows
have been reported in some Seyfert Galaxies \cite{krongold07},
often with sub-relativistic outflow velocities \cite{tombesi10}.
However, in this scenario a correlation between the global
wind covering factor and/or column density and the line-of-sight
column density could be expected. Such a correlation is not seen in the data.
While the complex nature of $C_s$ may conceal it, accepting this
interpretation would require {\it ad hoc} assumptions.
We
reject therefore this interpretation on astrophysical grounds.

\subsubsection{On the correlation between accretion rate and spectral shape}

A correlation between the accretion rate and the continuum spectral
shape was recently discovered in intensity-resolved {\it NuSTAR} spectra
of another nearby heavily obscured AGN, NGC~4945 \cite{puccetti14}.
Similar correlations were found by comparing single-epoch spectra
of large samples of Type 1 AGN \cite{shemmer08,risaliti09,brightman13}.
Assuming a standard conversion factor of 30 between the 2-10~keV and
the bolometric luminosity, $L_{bol}$ \cite{elvis94}
the absorption-corrected X-ray
AGN luminosity measured during the 2014-2015 campaign
(Table~\ref{table_mytorusbaseline}) converts
into an Eddington ratio
$\lambda_{Edd}$$\equiv$$L_{bol}/L_{Edd}$ between 1.9\%
and 3.0\%. The dynamic range is too small for a correlation against
the spectral index
to be measurable, given the typical statistical errors 
in each epoch (Table~\ref{table_mytorusbaseline}).
Conversely, the quality of time-resolved BAT spectra is
insufficient to reliably measure a spectral index that could correlated
with the larger flux dynamical range measured by {\it Swift}.
The data points are broadly consistent with
the locus in the $\Gamma$ vs. $\lambda_{Edd}$ plane occupied by
$z \approxlt 2$ AGN in the COSMOS and Extended {\it Chandra}
Deep Field South surveys \cite{brightman13}.

\subsection{X-ray constraints on the nature of the torus}

\subsubsection{X-ray time-averaged spectroscopy}
\label{sect_timeaveraged}

Fitting the multi-epoch X-ray spectra of the Markarian~3 campaign with models of toroidal
reprocessing yields constraints on the geometrical distribution of the
optically thick matter.
The Ikeda and the {\tt torus} models require large opening angles,
$\simeq$66$^{\circ}$ and $\simeq$58$^{\circ}$, respectively. The baseline Ikeda model requires a
grazing view along the rim of the torus ($\theta_{incl} \simeq$70$^{\circ}$).

Taking into account
the different geometries, the
estimates of the torus opening angle are in the broad agreement with the
opening angle assumed by {\tt mytorus} (60$^{\circ}$), making a comparison between
the results obtained with the Ikeda and the {\tt mytorus} models in the
``decoupled scenario'' meaningful in principle. However, there is a serious
tension between the Ikeda and {\tt mytorus} models as far as the
global column density out of the line-of-sight is concerned. The baseline
Ikeda model requires a Compton-thick column of $N_{H,sc}$$\simeq$$2.0 \times 10^{24}$~cm$^{-2}$,
a factor of $\simeq$2-2.5 larger than the line-of-sight
column density; the baseline {\tt mytorus} model, instead, prefers a
Compton-thin column ($N_{H.sc}$$\simeq$$1.2 \times 10^{23}$~cm$^{-2}$),
a factor of 7--9 lower than the line-of-sight column density \cite{yaqoob15}.
This tension is due to subtle differences in the curvature of the two
models (cf. Fig.~\ref{fig_models}) over an energy range still dominated
by the transmitted primary continuum in this source.
The small difference in fit quality between the two scenarios does
not allow us to prefer one solution on a purely statistical basis.
If the torus in Markarian~3 is seen through a line-of-sight grazing the upper
rim, however, one should expect the estimate of the global column density
to exceed the column measured in transmission if the torus is
thicker along the equatorial plane. On the basis
of this geometrical argument, we consider the solution based on the Ikeda
model more plausible on astrophysical grounds, and consider it as the reference
solution in the forthcoming discussion.

Finally, we comment on the Markarian~3 classification with respect to the
obscuration of its nucleus. Conventionally, the Compton-thin versus
Compton-thick classification is based on the measured column density
along the line-of-sight to the nucleus. We maintain this convention in
this paper. As $N_{H,abs}$ is always lower then $1.1 \times 10^{24}$~cm$^{-2}$
(Tab.~\ref{table_mytorusbaseline}), we conclude that Markarian~3 was
in a Compton-thin state during the 2014--2015 monitoring campaign, even
if the global column density estimated using the I09 model is
higher than the Compton-thick threshold.

\subsubsection{Line-of-sight $N_H$ variability: the 2014 occultation event}
\label{sect_2014}

\begin{figure*}
  \vspace{1.0cm}
  \includegraphics[height=130mm, angle=-90]{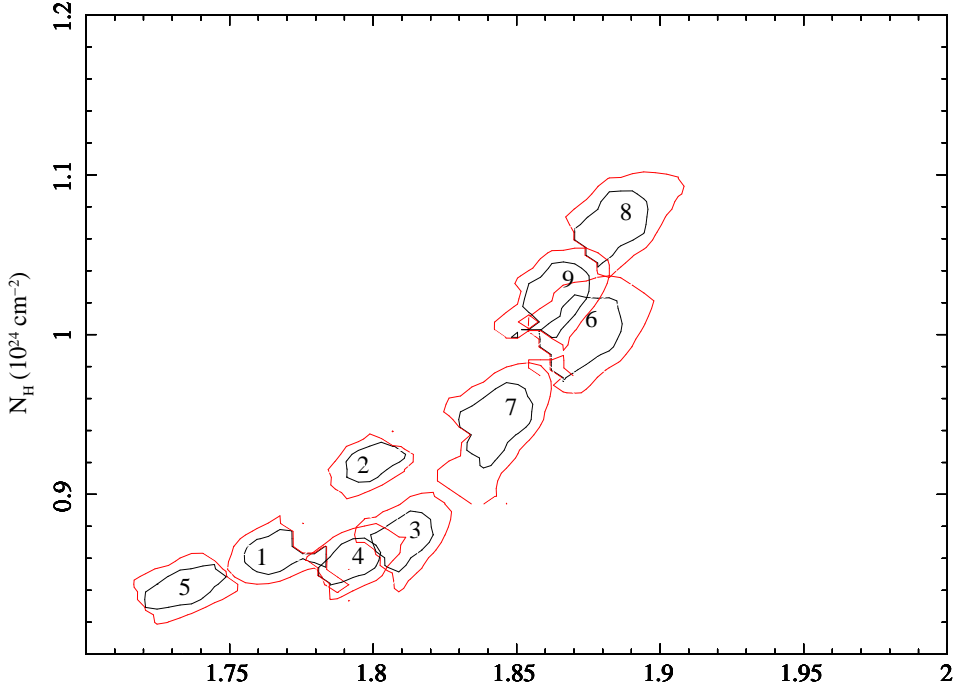}
  \caption{Iso-$\chi^2$ contours for the photon index versus the line-of-sight
    column density obtained from the Ikeda best-fit model applied to the
    spectra of the Markarian~3 2014--2015 monitoring campaign. The contours
    correspond to the 1-$\sigma$ and 90\% confidence levels for two interesting
    parameters. The {\it integer number labels} indicate the monitoring
    campaign epochs (cf. Tab.~\ref{tab_log}; 2014:1--5; 2015:6--9) and are placed at the
    locus corresponding to the best-fit values.
  }
  \label{fig_gammavsabsnh}
\end{figure*}

Additional constraints on the nature of the torus clouds along the
line-of-sight come from X-ray spectral variability.
In Fig.~\ref{fig_gammavsabsnh}
we show the iso-$\chi^2$ contours for the photon index $\Gamma$ versus
the line-of-sight
column density $N_{H,abs}$ obtained from the Ikeda baseline model applied to the
spectra of the Markarian~3 2014--2015 monitoring campaign.
Using the two-dimensional contours allows us to
estimate the dynamical range and time-scale of the line-of-sight
column density variability, taking properly into account
that this parameter is degenerate with the continuum shape in the
spectral fitting procedure\footnote{The Spearman $\rho$
correlation coefficient (probability) is 0.97 (99.998\%), 0.90 (96\%), and
0.80 (80\%) for the full campaign,
the 2014 and the 2015 observations, respectively.}.
We obtain similar results
if we plot the column density against the power-law normalization,
or if we use the results obtained with the {\tt mytorus} model.
The column density
shows two distinct variability patterns during the two sub-epochs
of our monitoring campaign. We discuss these patterns separately in this,
and in the following sub-section.

During the
2014 observations the column density remains constant within the
statistical uncertainty, with the exception of an increase by
($\Delta N_H$=$(4.9\pm1.4) \times 10^{22}$~cm$^{-2}$) between the first
(September 7) and second (September 14) observations, followed
by a recovery to the original obscuration level
on October 1. The most straightforward interpretation
of this finding is an occultation event by a single additional cloud.
Assuming that the total line-of-sight column density is due to the
superposition of a system of identical (in $N_H$) clouds, the
total number of line-of-sight clouds can be constrained to
$17\pm5$.

From the time-scale of this event $t_{occ} \simeq 30$~days,
we can estimate
the distance of the occulting cloud from the
X-ray source, $R$, under the assumption that
its main velocity component is due to the orbital velocity that can
be well approximated by the Keplerian velocity at $R$:
\begin{equation}
  v_k = \sqrt{\frac{G M_{BH}}{R}}
\end{equation}
where $G$ is the gravitational constant.
Assuming spherical symmetry and homogeneity, we can write an expression
for $R$ assuming that we can approximate the (unknown) size of the
cloud $s \equiv v_k/t_{occ}$ as $\simeq$$N_H/n$, where $n$ is the
particle density \cite{lamer03,risaliti07,svoboda15}:
\begin{equation}
  R \simeq 1.6 \times 10^{14} \, M_{Mkn3, BH} \, t^2_{Mkn3, occ} \, \Delta N^{-2}_{H,Mkn3} \, n^2_6 \; cm
  \label{eq_r}
\end{equation}
where the ``Mkn3'' subscript indicates that we have expressed the
corresponding quantity in units corresponding to the occultation
event in Markarian~3.
To close the system we would need an estimate of the electron density
of the cloud. We can make the additional assumption that the average
ionization status of the occulting cloud is the same of the gas
responsible for the bulk of the Fe K$_{\alpha}$ emission line. The
centroid energy as measured in the {\it Chandra}/HETG spectrum
is $6.396\pm0.008$~keV, while the energy of the photo-absorption edge
in the same data is constrained to be $7.13\pm0.03$~keV.
These measurements correspond to a Fe ionization state
Fe$\le${\sc v} \cite{kallman04}, or $\log(\xi) \approxlt -1.5$,
where $\xi$ is the ionization parameter $\equiv \frac{L}{n R^2}$ in
c.g.s. units,
and $L$ is the illuminating luminosity in the energy range
13.6~eV--13.6~keV ($\sim$1.1$\times 10^{44}$~erg~s$^{-1}$ in Markarian~3).
Eq.~\ref{eq_r} would therefore imply a cloud at a distance
$R \approxgt$4$\times 10^{18}$~cm, with a density $n \approxlt 10^4$~cm$^{-3}$.

\subsubsection{Line-of-sight $N_H$ variability in 2015}
\label{sect_2015}

During the 2015 observations the
line-of-sight column density variability pattern
is more complex, and no clear individual occultation event can be
identified.
The largest $N_H$ variation occurred between March 23 and April 5
($\Delta N_H$=$(1.3\pm0.3) \times 10^{23}$~cm$^{-2}$). A
$\Delta N_H$=$(5\pm 3) \times 10^{22}$~cm$^{-2}$ was
measured between the two observations with the shortest
separation (three
days), still significant at the 90\% confidence level for two
interesting parameters. We compared the 2015 $N_H$ measurements
with the expected Poissonian distribution due to an average
number $N_c$ of
clouds with the same column density $N_{H,c}$, and searched for the $N_c$
maximizing the Kolmogornov-Smirnov
probability that the observed measurements are extracted from
that distribution. The maximum probability occurs for
$N_{H,c}$=$(3.2 \pm 0.4) \times 10^{22}$~cm$^{-2}$, or
$N_c$=$(32 \pm 4)$. While the assumptions behind this estimate
are admittedly strong, as also confirmed by the shallow
K-S probability ($\simeq$70\%), the
resulting number of line-of-sight clouds is within a factor of two that
derived from the single cloud occultation event in 2014.

\subsection{Comparing the torus view in X-rays and IR}
\label{sect_disctorus}

These results can be compared to the constraints on the torus structure
derived from
IR observations. The most complete study to-date
is based on Gemini/Michelle spectra at a resolution of 200~pc
\cite{sales12}.
Clumpy torus models \cite{nenkova08} require an opening angle of 50$^{\circ}$,
seen
at an inclination between 53$^{\circ}$ and 70$^{\circ}$ (best-fit 66$^{\circ}$).
The inclination angles
are in excellent agreement with those derived from X-ray spectroscopy.
However, we stress that these results are obtained using a model with
very different assumptions on the torus structure:
clumpy in IR, homogeneous
in X-ray models. Indeed, the mere detection of an occultation event
during the 2014 campaign, and the line-of-sight absorbing column density
variability observed during the 2015 campaign are clear indications
of a torus clumpy structure in Markarian~3.

In the Nenkova et al. (2008) model the torus is parameterized as a system of discrete clouds.
Such a system is characterized
by the average number of clouds along the torus equatorial plane; the visual optical depth of each
cloud; the radial extension and profile of the cloud distributions, besides the inclination
and the opening angles. The total column density of the IR
clumpy torus in Markarian~3
was constrained to be $(5 \pm 3)$$\times$$10^{23}$~cm$^{-2}$
\cite{sales12}, {\it
i.e.} a factor of 4 lower than
the global column in X-rays in the baseline Ikeda model,
and consistent with the {\tt mytorus} baseline model.
The average number of clouds in the equatorial plane ($6\pm^3_1$)
corresponds to
$4 \pm^2_1$ along the estimated inclination angle to the torus (66$^{\circ}$).
Once again, this is a factor of about 3 lower than the
number of line-of-sight clouds estimated through the study of
the X-ray absorption variability pattern.
Each of the IR-emitting clouds has an optical depth of $76 \pm 12$~mag,
corresponding to $(1.7 \pm 0.3) \times 10^{23}$~cm$^{-2}$, i.e. a factor
of about 4 larger.

Comparing the properties of the IR-emitting dust torus with that
of the X-ray absorber/scatterer 
may indicate that the X-ray absorbing gas is
largely within the dust sublimation radius.
The mean dust sublimation radius for Silicate grains can
be expressed as
$r_{sub}$$\sim$$L_{46}^{1/2} [T/1500 K]^{2.6} f(\phi)$~pc, where $L_{46}$ is the AGN
bolometric luminosity in units of 10$^{46}$~erg~s$^{-1}$ and $f(\phi)$
is a geometrical factor equal to
1 for an isotropic source \cite{barvainis87}.
The absorption-corrected 2--10~keV AGN luminosity measured during the 2014--2015
campaign ranged between 3.5 and 5.4$\times$10$^{43}$~erg~s$^{-1}$
(the typical error on each
measurement is between 10\% and 20\%). Assuming
a bolometric correction of 30 \cite{elvis94,vasudevan09},
the dust sublimation radius is $\sim \times$10$^{17}$~cm
(0.4~pc).

However, such a conclusion relies on an admittedly crude understanding of
the torus geometry. Other results presented in this paper point to
a more complex and extended structure of the gaseous phase,
either in the radial or in the vertical
direction. If the occultation event in 2014 is due to
a single cloud crossing our line-of-sight, it
should be located at a distance of about 10~pc from the central engine
(\$~\ref{sect_2014}).
IR spectroscopy suggests that torus
extends to a radius of about 7~pc \cite{sales12}. Hence, such
a cloud could belong to the torus outskirts. Another independent
constraint on the location of the reprocessing gas
comes from the width of the Fe K$_{\alpha}$ fluorescent line.
Following Gandhi et al. (2015), we estimate
the emission radius of this line, $R_{Fe}$, as:
$$
R_{Fe} \equiv \frac{G M_{BH}}{v^2}
$$
where $v$=$\sqrt{3}/2 v_{FWHM}$. Using the line
width measured in the {\it Chandra}/HETG spectrum (\$~\ref{sect_hetg}),
and applying the 2.3548 Gaussian conversion factor between standard
deviation
and Full Width Half Maximum, $R_{Fe}$$\simeq$$8\pm3$~pc
(we assumed here
a fiducial 25\% error on the black hole mass
in this obscured AGN). This estimate is a factor of about 20 larger than
the dust sublimation radius, $R_{dust}$. This is in contrast
to most of the AGN
in the Gandhi et al. (2015) sample, where $R_{Fe}$ is comparable or
lower than $R_{dust}$. We interpret this discrepancy in terms
of line-of-sight orientation with respect to the torus axis. Most
of the objects in the Gandhi et al. (2015) sample are Seyfert~1--1.9.
They are not heavily obscured in X-rays, and are
probably seen along small inclination
angles. In Markarian~3, instead, the large inclination angle may prevent
us from directly viewing the innermost regions of the torus in line
emission. The
bulk of the observed Fe line profile would therefore come from
a more extended region, visible to us above the torus rim.

Such an
interpretation could be validated by better Fe iron line data, allowing
a more accurate decomposition of the line
profile. This is impossible with the {\it Chandra}/HETG, but will
be soon possible with with the micro-calorimeter on board {\it ASTRO-H}
\cite{mitsuda12}, thanks to its superb resolution and larger
collecting area at the Fe line energies.

\subsection{An ionized absorber in Markarian~3?}
\label{sect_ionizedabsorber}

The baseline models during the 2014--2015 monitoring campaign require an
absorption line around 7~keV (source frame) during most of the epochs.
Its centroid energy is not very well constrained, but it is
consistent with resonant absorption from H-like iron. This feature
has never been reported in previous X-ray spectroscopic analysis of Markarian~3.
Its presence is confirmed in the higher-resolution
HETG spectrum,
where the single profile can be fit with complex structure
composed of a recombination Fe{\sc xxv} emission line, and two absorption lines
whose centroid energies are
consistent with resonant Fe{\sc xxv} and Fe{\sc xxvi} absorption.
The intensity of the absorption lines in the {\it Chandra}/HETG spectrum
is lower than observed in 2014, and consistent with that observed in 2015.

This feature could be the signature of a highly ionized absorption
component along the line-of-sight to the AGN, possibly transient.
A full characterization of the physical properties of this
absorbing system is hampered by the quality of the data,
insufficient, for instance, to assess if this absorption system is static or outflowing.
From the theoretical standpoint,
outflows are expected to arise in a disk atmosphere across or beyond
the dust sublimation radius \cite{czerny11}.
Curve of growth analysis \cite{bianchi05a,risaliti05} requires
a column density $\approxlt$10$^{22}$~cm$^{-2}$ for the shallow
absorption state measured in 2015,
if the absorbing gas
sees a Seyfert-like Spectral Energy Distribution, and the turbulent velocity
(unconstrained in our data) is lower than 500 km~s$^{-1}$.
This means that the absorber would be comparable to the ``warm absorbers''
commonly observed in Seyfert Galaxies \cite{mckernan07,laha14},
and probably launched at distances consistent with the innermost regions
of the torus \cite{blustin05}.
On the other hand, the
column density estimate
increases by at least one order of magnitude for the deeper features
measured in 2014. Still, the constraints on the outflow
velocity prevents the absorbing system in Markarian~3 from being
identified as an Ultra Fast Outflow \cite{tombesi10}.
We can compare it rather with the outflow detected in another
(typically; Braito et al. 2014)
heavily obscured AGN: NGC~1365 \cite{risaliti05}, where a system
of four transitions were detected, corresponding to K$_{\alpha}$ and
K$_{\beta}$ resonant absorption by He- and H-like iron. In Markarian~3
the EW is at least one order of magnitude lower than in NGC~1365, though.

The discovery of this highly ionized absorbing system is intriguing. It
shows the not-often-detected simultaneous presence of cold optically thick
obscuration and of an optically thin warm absorber in the same active galaxy.
This
provides indirect support for an origin
of the ionized absorber as clouds evaporated
at the outer surface of the torus \cite{kartje96,blustin05}, and possibly
accelerated by the radiation pressure
due to the central AGN emission leaking through
the patchy absorber. The existing anti-correlation between the
line-of-sight column density and the warm absorber feature EW
(cf. Table~\ref{table_mytorusbaseline}) lends support
to this interpretation. Alternatively, the gas could be ionized along an
absorption-free line-of-sight to the AGN, and fall back towards the
equatorial plane in a failed jet or outflow \cite{ghisellini04,czerny11}.
However, one should bear in mind that Markarian~3 does host a radio
jet \cite{pedlar84}.

Future high resolution measurements in the
energy range where the Fe atomic transitions occur, such as those
possible with the micro-calorimeter on board {\it ASTRO-H}
\cite{mitsuda12} will be crucial to properly characterize the
physical properties of this absorbing system, and perform the
required time-resolved
diagnostics.

\section{Summary and conclusions}
\label{sect_summary}

We report in this paper the results of an X-ray monitoring campaign
of the heavily obscured Seyfert Galaxy Markarian~3,
carried out between the Fall of 2014 and the Spring of 2015
with {\it NuSTAR}, {\it Suzaku} and {\it XMM-Newton}. The campaign consisted
of nine epochs, covering a wide range of
time separations (and therefore
of potential source variability time-scales) from
about 3 days to 7 months. The
campaign aimed at constraining the properties of the
obscuring and reflecting material in this bright and
highly variable AGN
by comparing the X-ray spectra with geometrically motivated
models of Compton scattering by optically thick matter in
the AGN environs, as well as by analyzing its extreme
spectral variability on time-scales much shorter than studied so
far (Iwasawa et al. 1993, G12).

The main results can be summarized as follows:

\begin{itemize}

\item The hard X-ray spectrum of Markarian~3 is variable on
  all the time scales probed by our campaign, down to the shortest separation
  between consecutive observations (4~days;
  cf. Sect.~\ref{sect_dataanalysis}).

\item We disprove the claim originally made by G12 (and already
  criticized by Yaqoob et al. 2015) that the
  X-ray spectral variability of the continuum below 10~keV can be
  used to constrain the properties of the optically
  thick reprocessor in this
  object.
  This variability is due to a combination of a variation of the
  primary continuum (cf. Sect.~\ref{sect_originvariability}) and
  of the intervening line-of-sight absorber column density
  (cf. Sect.~\ref{sect_2014} and \ref{sect_2015})

\item If arranged in a spherical-toroidal geometry as assumed by the
  Ikeda (I09) model, the
  Compton scattering torus has an opening angle $\simeq$66$^{\circ}$, and is seen
  at a grazing angle through its upper rim
  (inclination angle $\simeq$70$^{\circ}$).
  The global average column density is $\sim 2 \times 10^{24}$~cm$^{-2}$,
  keeping Markarian~3 in the rank of Compton-thick AGN even if the
  line-of-sight column density measured during the monitoring campaign
  is in the Compton-thin range (0.8--1.1$\times 10^{24}$~cm$^{-2}$;
  cf. Sect.~\ref{sect_timeaveraged}).

\item We report the discovery of an increase of the line-of-sight column
  density during the 2014 observation, followed by a subsequent recovery to
  the pre-rise level. If due to an occultation event
  by a single cloud belonging to a system of clouds sharing the same
  column density, this event allows us to constrain their number
  ($17 \pm 5$) and individual column density,
  [$(4.9 \pm 1.5) \times 10^{22}$~cm$^{-2}$]
  (cf. Sect.~\ref{sect_2014}).

\item While we cannot identify an occultation event during the 2015
  campaign, the observed variability pattern of the line-of-sight
  column density is consistent within a factor of two
  with the geometrical and physical
  properties of the absorber as derived from the 2014 event
  (cf. Sect.~\ref{sect_2015}).

\item The combination of the two previous pieces of evidence lends clear
  support to the clumpy nature of the torus in Markarian~3, as also
  indicated by the difference between the line-of-sight and
  global column density inferred by all models discussed in this paper.

\item The comparison between the derived properties of the obscuring gas
  and the properties of the IR-emitting dust
  \cite{sales12}
  suggests that at least two thirds
  of the X-ray obscuring gas volume might be
  located within the dust sublimation radius.
  However, the most dynamical clouds -- such as those responsible for the
  occultation event in 2014 -- are probably located on a larger
  scale, in the outskirts of the dusty torus
  (cf. Sect.~\ref{sect_disctorus}). While the derived geometries of
  the IR- and X-ray material are similar, we stress
  that they were derived using different
  assumptions on the torus structure -- clumpy in IR, homogeneous
  in X-ray models

\item We report the discovery of ionized absorber, characterized
  by variable resonant absorption lines due to He- and H-like iron. Markarian~3
  is, to the best of our knowledge, the second object after NGC~1365
  where an ionized absorber has been detected alongside heavy
  X-ray obscuration. This discovery lends support to the idea that
  moderate column density absorbers could be due to clouds evaporated
  at the outer surface of the torus, possibly
  accelerated by the radiation pressure
  due to the central AGN emission leaking through
  the patchy absorber \cite{czerny11}
  (cf. Sect.~\ref{sect_ionizedabsorber}). The combination of iron
  absorption and emission features 
  makes of Markarian~3 an ideal target to study the relation between torus
  reprocessing and the onset of ionized outflows with future
  high resolution instruments such
  as the micro-calorimeter on board {\it ASTRO-H} \cite{mitsuda12}.
  
\end{itemize}

\section*{Acknowledgments}
This work made use of data from the NuSTAR
mission, a project led by the California Institute of Technology,
managed by the Jet Propulsion Laboratory, and funded
by the National Aeronautics and Space Administration. We
thank the NuSTAR Operations, Software and Calibration
teams for support with the execution and analysis of these
observations. This research has made use of the NuSTAR
Data Analysis Software ({\tt NuSTARDAS}) jointly developed
by the ASI Science Data Center (ASDC, Italy) and the California
Institute of Technology (USA).
JS acknowledges support from the grant LH14049 and the
Project 14-20970P of the Grant Agency of the Czech Republic.
FEB acknowledges support from CONICYT-Chile (Basal-CATA PFB-06/2007, FONDECYT Regular 1141218, "EMBIGGEN" Anillo ACT1101), and the Ministry of Economy, Development, and Tourism's Millennium Science Initiative through grant IC120009, awarded to The Millennium Institute of Astrophysics, MAS.
The authors are grateful to an anonymous referee, whose
accurate and detailed report greatly improved the clarity of the paper.

\label{lastpage}

\end{document}